\documentclass[aps]{revtex4-1}
\usepackage{graphicx}
\usepackage{natbib}
\usepackage[colorlinks,urlcolor=blue,citecolor=blue,linkcolor=blue]{hyperref}
\usepackage{amsmath}
\usepackage{subcaption}
\usepackage{gensymb}

\usepackage{graphicx}
\usepackage{wrapfig}
\usepackage{natbib}
\usepackage{url}
\usepackage{enumitem}

\usepackage{graphics}

\newcommand{\be}{\begin{equation}}
\newcommand{\ee}{\end{equation}}
\newcommand{\nn}{\mbox{} \nonumber \\ \mbox{} }
\newcommand{\ba}{\begin{eqnarray}}
\newcommand{\ea}{\end{eqnarray}}
\newcommand{\om}{\omega}
\newcommand{\Alfven}{Alfv\'{e}n }

\newcommand{\B}{{\bf B}}

\newcommand{\Bf}{{magnetic field}}

\newcommand{\Ef}{{electric  field}}

\newcommand{\NSs}{{neutron stars}}
\newcommand{\EM}{electromagnetic}

\newcommand{\mss}{magnetospheres}

\begin{document}

\title{Powerful parametric instability of \Alfven waves in astrophysical pair plasma }

\author{ Maxim Lyutikov \\
Department of Physics and Astronomy, Purdue University, \\
 525 Northwestern Avenue,
West Lafayette, IN
47907-2036 }

\begin{abstract}
We demonstrate that    in  highly   magnetized pair plasmas,   nonlinear   \Alfven waves with wave-number $k \leq  k_0 = \omega_p^2 /(\delta \om_B)$   ($\delta =( \delta B)/B_0$ are relative fluctuations of the \Bf)
experience powerful  modulational  instability. In the two-fluid approximation, we develop an analytic set-up 
 for circularly polarized (CP)  \Alfven mode in its frame  (where the  initial configuration  is stationary; it is moving with relativistic, amplitude-dependent \Alfven velocity $v_A (\sigma, \delta )  $, while both charges experience different, amplitude-dependent,  synchrotron  gyration).  PIC simulations using EPOCH code demonstrate that for \Alfven  waves with $k$ near $k_0$,
large,   parametrically-driven   density fluctuations develop, and lead to fast modulational instability. Charge separation effects,  for a  CP wave  in magnetized  pair plasma,  might be temporarily important;  on longer time-scales the density fluctuations are charge neutral and in symmetric pair  plasma quickly  grow to large amplitudes.  In highly magnetized plasma,  $\sigma \gg 1$,
 high frequency modes $k / k_0 \sim (2-3 ) \times \sigma \gg 1 $ are quickly generated; for  smaller plasma magnetization, the dominant mode is at the  Bragg's condition $k = 2 k_0$.   Long term behavior of CP and LP modes is similar.  We discuss application of the results to the physics of  Fast Radio Bursts generated/propagating  in the  \mss\ of magnetars.
\end{abstract}

\maketitle 
   
   \section{Introduction}
   
   Dynamics of nonlinear waves in plasmas is a classical problem in plasma physics  \citep{AkhiezerPolovin,1975OISNP...1.....A,1974JPlPh..12..297C} that  recently became important for astrophysical Fast Radio Bursts \citep[FRBs][]{2007Sci...318..777L,2022A&ARv..30....2P,2019ARA&A..57..417C}.
   
     The present work,
     following  on \citep{2025arXiv250906230T,2025arXiv250920594L,2025arXiv250917245L},  investigates modulational instability of nonlinear waves in astrophysical pair plasmas, taking into account the guide \Bf, see Appendix \ref{modulational} for a discussion of previous works. \citep[In Ref.][the initial set-up is not  nonlinear \Alfven waves, as fields and currents do not match to a specific  eigenmode of a medium.]{2021ApJ...915..101L}
     

In this work  we explore    parametric instability of \Alfven waves in the nonlinear  regime when the fluctuating field $\delta B$  is comparable  than the guide field $B_0$.  As we demonstrate, the waves are subject to a powerful modulational instability, specific to pair plasma.

     \section{Relativistically nonlinear \Alfven waves in   \Alfven frame}
     
     \subsection{General relations}
     
    Relativistically nonlinear circularly polarized (CP)  \Alfven\  waves, with $\delta_A  \equiv \delta B /B_0$, may be considered in a two-fluid approximation in a fully analytical form \citep{2025arXiv250917245L}. For  sub-luminal  phase velocity,  there is a frame,  the \Alfven\  frame,  where \Ef\ is zero, while $\nabla \times \B = \ {\bf j}$ (we use abbreviated notations, with factor $4 \pi$ absorbed into definition of a charge; resulting relation look SI-like, with $\epsilon_0$  and $\mu_0$ set to unity; in all cases the momenta are normalized to $m_e c$).  In the \Alfven  frame, the  plasma is moving along the guide field with velocity $v_A (\sigma_A ,    \delta_A, k_A  )$, to be determined self-consistently; $v_A$   depends on  the plasma magnetization  parameter  $\sigma_A$ (as measured in the  \Alfven  frame -  it is different from the  magnetization  parameter $\sigma_0$ in the plasma frame  $\sigma_0$ by the Lorentz transformation of the density, see Eq. (\ref{sigma0}))
    \be
    \sigma_A = \frac{\om_B^2}{\om_{p,A} ^2}
    \label{sigmaA}
   \ee
   (plasma frequency $\om_{p,A}$ is  measured in   \Alfven frame and does not include any relativistic contributions to the effective mass, only density).
   
   A stationary CP wave in the \Alfven frame is  described by 
   \ba && 
   {\bf e}_B = \{ 0 , \sin( k_A x) , \cos ( k_A x) \}
   \nn &&
   \B = \left(  {\bf e} _x  + {\bf e}_B \delta_A  \right)  B_0 
   \ea
   (wave number $k_A$ and wave intensity  $ \delta_A$   are measured in   \Alfven frame, different from the plasma frame),

   For pair plasma, as a basic set-up, we can assume that  parallel {\it velocities}  $v_A$ of both components are equal. This is not obvious from the start:  for a CP wave packet coming into  pair plasma,  the ponderomotive force is different on electrons and positrons. But  the resulting parallel electric field then quickly accelerates the slower component/decelerates the faster one. As a result both specie move with eh same parallel velocity  (we verified this   via direct  PIC simulations).
   
   Parallel velocities are equal, but transverse velocities/momenta of two species  are different since one of the species  is  resonant. There are many possible parametrization of the set-up, we  use the following one:  given are parallel velocity $v_A$ and transverse velocities 
   \ba && 
   v_A = \frac{p_A}{\sqrt{1+p_A^2}}
   \nn &&
   v_{0, \pm} = \frac{p_{0, \pm}} { \sqrt{  1+ p_{0, \pm}^2}  \sqrt{1+p_A^2}} 
   \ea
   
   Importantly, $p_A$ is NOT the parallel momentum, it's just a parameter related to the parallel velocity. 
   We also note the relation
   \ba && 
   \gamma_\pm = \gamma_A   \gamma_{0, \pm}
   \nn &&
    \gamma_A  = 1 /\sqrt{1- v_A^2} = \sqrt{1+p_A^2}
    \nn && 
     \gamma_{0, \pm} =  \sqrt{  1+ p_{0, \pm}^2}  
     \ea

   There are three resulting equations: the current balance, and force balance
   \ba &&
   {\bf j} = \nabla \times \B = n_0 ({\bf v}_+ - {\bf v}_-)
   \nn && 
   v_A \partial_A {\bf p}_\pm = \pm {\bf v}_\pm \times {\B}_0
   \ea
   
  Scaling $k_A = {\cal K}_A \om_B$,  We find 
   \ba &&
      \delta_A {\cal K}_A \sigma_A  = \left(   \frac{ p_{0, p}} {\gamma_{0, p}} -  \frac{ p_{0, e}} {\gamma_{0, e}} \right) \frac{1}{\gamma_A}
   \nn &&
      \delta_A= \left(  \frac{1}{\gamma_{0, e} p_A}  + {\cal K}_A   \right) p_{0, e}
   \nn &&
      \delta_A=  \left(  \frac{1}{\gamma_{0, p} p_A}  - {\cal K}_A   \right)  p_{0, p}
   \label{main}
\ea
For  given plasma and wave parameters $\sigma_A$, $ {\cal K}_A$ and $   \delta_A$, 
equations (\ref{main}) determine momenta of particles  $p_{0, \pm}$ and the  \Alfven momentum $p_A$.  All quantities are positive, as the signs of charges and the sense of circular polarization are explicitly taken into account. By our choice of polarization, positrons can be in resonance, when
\be
 {\cal K}_A =  \frac{1}{\gamma_{0, p} p_A}  \to  p_{0, p} =  \frac{\sqrt{1 -  {\cal K}_A^2 p_A^2}}{{\cal K}_A p_A }
 \ee

The set-up (\ref{main}) is highly specific, as the  stationary nonlinear \EM\ fields are exactly balanced by $e^\pm$ currents in the wave frame.

       \subsection{Limiting cases}
       
       The  set-up of a  relativistic nonlinear \Alfven wave in its frame is somewhat unusual. As a check, 
     in the limit $k_A \to 0$, we find the \Alfven\ momentum
 \be 
   p_A =\frac{\sqrt{\sigma_A } \sqrt{\sqrt{ 16+ \left(1+   \delta_A ^2\right)^2 \sigma_A ^2}+\left(1-   \delta_A
   ^2\right) \sigma_A }}{\sqrt{2} \sqrt{4+    \delta_A ^2 \sigma_A ^2}}   \to \frac{\sqrt{\sigma_A  \left(\sqrt{16+\sigma_A ^2}+\sigma_A \right)}}{2 \sqrt{2}}
   \label{pAk0}
   \ee
   The corresponding  transverse momenta are 
   \ba &&  
   p_{e,p} = p_0 \mp    \Delta   {\cal K}_A
   \nn &&
   p_0 = \frac{   \delta_A  p_A}{\sqrt{1-   \delta_A ^2 p_A^2}}
   \nn &&
       \Delta= \frac{ \sqrt{1+p_A^2}  \sigma_A    \delta_A  }{2  \left(1-   \delta_A ^2
   p_A^2\right){}^{3/2}}   
   \ea
   These are relations for fully nonlinear  CP \Alfven wave in the limit $k_A \to 0$ as expressed in terms of parameters measured in the \Alfven frame.  In expression for $p_A$ we also give a linear limit $   \delta_A \to 0$.
   
   To recover a more familiar form, we need to substitute 
   \be
   \sigma_A \to \sigma_0 /\gamma_A,
   \label{sigma0} 
   \ee
    where $ \sigma_0$ is the sigma-parameter defined in terms of lab quantities. Density in the Alfven frame is higher by $\gamma_A$.  We then recover 
   \be
   p_A =\frac{\sqrt{\sigma_0 \left(\sqrt{16+    \delta_A ^4 \sigma_0^2}-   \delta_A ^2 \sigma
   _0\right)}}{2 \sqrt{2}} \to \sqrt{\sigma_0/2}
   \label{vA}
   \ee
   (factor of $2 $ comes from the fact that we define sigma-parameter with respect to each component separately). 
   
   Eq. (\ref{vA}) gives \Alfven momentum for nonlinear CP \Alfven wave in plasma with $\sigma_0$ in the limit $k\to 0$, as measured in the lab frame. Such wave always exists  (but only in the limit $k \to 0$, see below).

    In the linear regime $   \delta_A\ll 1$ (but $p_A$ may still be large in highly magnetized plasma)
    \ba &&
    p_{0, p} = \frac{p_A}{1-  {\cal K}_A  p_A}    \delta_A
    \nn &&
     p_{0, e} = \frac{p_A}{1+  {\cal K}_A  p_A}    \delta_A
     \ea
     \be
     \gamma_A \sigma_A =  2 \frac{ p_A^2}{1- {\cal K}_A ^2  p_A^2}
     \label{disp1}
     \ee
     Equation (\ref{disp1}) is in fact a conventional  linear \Alfven velocity: in lab frame
    \ba &&
v_A =     \frac{\om_0}{k_0} = \sqrt{  1- \frac{ 2 \om_{p,0}^2}{\om_0^2- \om_B^2}}
\nn &&
\gamma_A =  \sqrt{1+ \frac{\om_B^2 - \om_0^2}{ 2 \om_{p,0}^2}}
\ea
Defining plasma frequency  $ \omega_{p,0} = \sqrt{4 \pi n e^2/m}$ (without $\gamma_A$ in particle mass), we recover (\ref{disp1}) with Lorentz-transformed wave vector
\be
k_A = \sqrt{2} \frac { \om_0 \om_{p,0} }{\sqrt{ \om_B^2 - \om_0^2}}
\label{kkA} 
\ee
Since  in the lab frame $0< \om_0 <\om_B$,  in the \Alfven frame $0<k_A < \infty$.

      \subsection{"End of  dispersion"}

For any finite wave  amplitude  $   \delta_A$,  there is a terminal wave number  ${\cal K}_A^\ast$ beyond which \Alfven waves do not exist \citep[as shown in][]{2025arXiv250917245L}.
When expressed in terms of the \Alfven frame parameters,  nonlinear  \Alfven\ waves  exist for 
\be
{\cal K}_A ^{(max)} \leq  \frac{1}{   \delta_A \sigma_A}
\ee


 \begin{figure}
 \includegraphics[width=.99\linewidth]{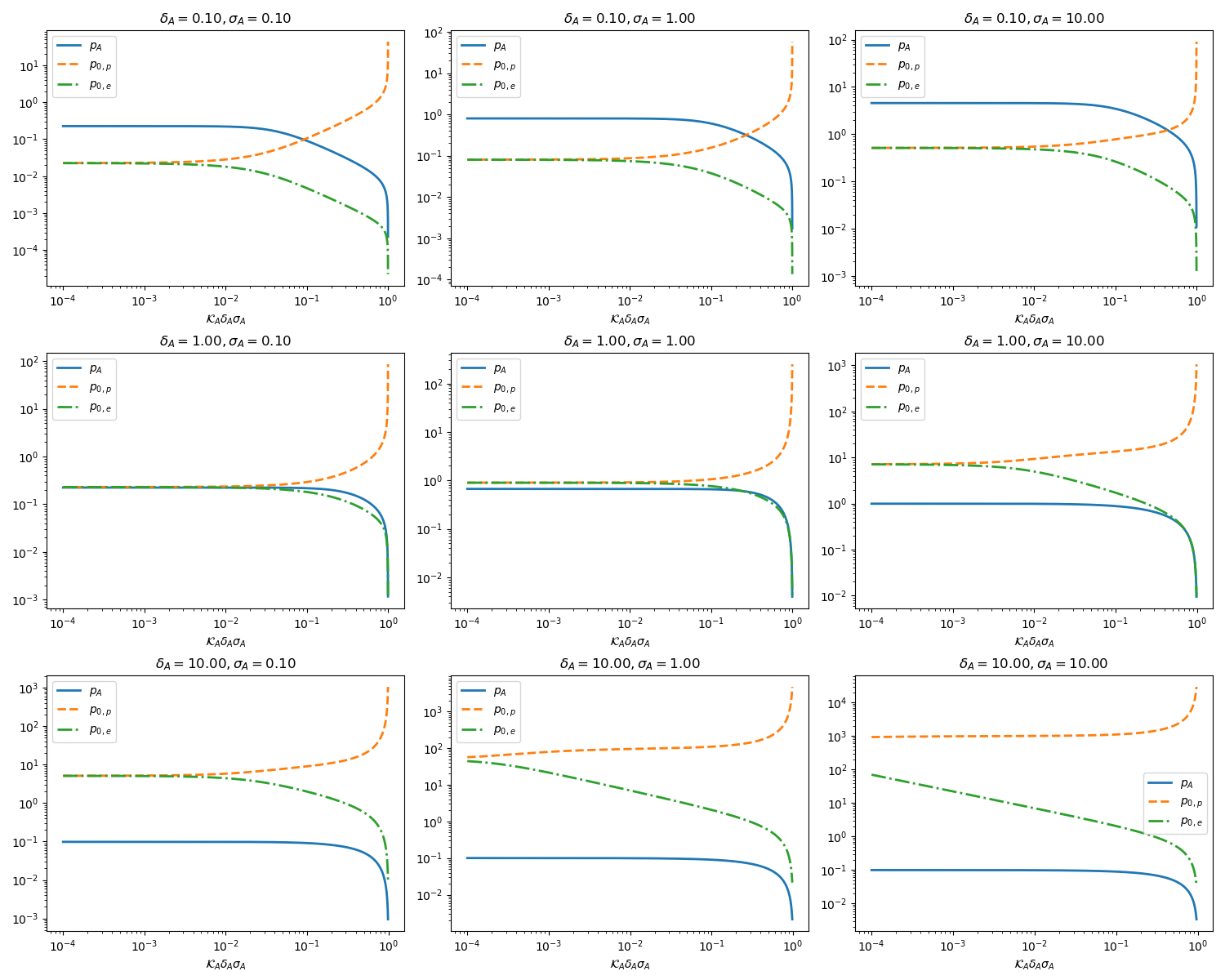}
  \caption{ Alfven momentum $p_A$, and pairs' momenta $p_{0,\pm}$ (solution of Eqns. (\ref{main})) as function of $ {\cal K}_A$ scaled to $1/(  \delta_A \sigma_A)$. }
\label{Roots_Tracking}
\end{figure}

In
Fig. \ref{Roots_Tracking}  we plot \Alfven momentum $p_A$ and particles' momenta $p_{0, \pm} $ as a function of wave number.
We conclude:  (i) solutions for nonlinear \Alfven waves always exist in the limit ${\cal K}_A \to 0 $;  (ii) for any $   \delta_A$ and $\sigma_A$ there is a characteristic ${\cal K}_A =  1/(    \delta_A \sigma_A)$ where $p_A$ becomes zero - beyond this point (larger $k_A, \delta_A$ and/or $\sigma$)  \Alfven waves do not exist. 

Close to the terminal point,
\ba &&
 p_A \approx  \frac{1 - \mathcal{K}_A \delta_A \sigma_A}{\delta_A} 
 \nn &&
p_{0,e} \approx 1 - \mathcal{K}_A \delta_A \sigma_A 
\nn && 
 p_{0,p} \approx  \frac{\delta_A^2 \sigma_A}{1 - \mathcal{K}_A \delta_A \sigma_A} 
 \ea

\citep[Connection to the lab-frame quantities are highly  non-trivial, as the phase speed of \Alfven waves depends on the wave amplitude; also in  Ref. ][results are presented in terms of $\eta _w= E_w/B_0 = v_{\rm ph} \delta _{0}$  ]{2025arXiv250917245L}

In what follows we investigate the modulational instability of \Alfven waves in the propagating regime ${\cal K}_A \leq {\cal K}_A ^{(max)}$.
  
  \section{Results of simulations}

   \subsection{Simulations set-up and overall conclusion}
   
   The simulations were performed using the EPOCH code \cite{Arber:2015hc}.  Our typical run has $n_p =100$ particles per cell and $n_x=100$ cells per period. Run  times vary from $30 c/\lambda$ to $500 c/\lambda$.

   A  challenging part is setting up  nonlinear \Alfven waves in their  own frame: particles' velocities and \EM\ field must match. Two important points are: 
 (i)   in EPOCH one needs to specify initial momenta of particles. Both types has the same parallel velocity, but since perpendicular energies are different, the parallel  momenta, for the same velocity,   are different   as well; (ii) the initial particle velocities {\it  are}  treated as a current, so that at $t=0$, the code  evaluates $\mathbf{E}^{1} = \mathbf{E}^{0} + \Delta t( \nabla \times \mathbf{B}^0 - \mu_0 \mathbf{J}^{1/2})$; a  half-step mismatch between current and \EM\ fields can be made small enough for sufficiently high resolution $n_x$; (iii) Fully nonlinear set-up  is possible only for CP. 
  We verified temporal stationarity (correct initialization)  on $\sim $ few periods. 

As the modulation occurs on a scale of a fraction of the initial wavelength, the most revealing simulations involve initial box of one $\lambda$. Longer wave trains require larger resolution, and are generally consistent, Fig. \ref{compare-resolution}.
Since the system is nonlinear, only a  number of relations/correlations  can established by running codes with different parameters.    

   As a summary:
\begin{itemize}
  \item  Large density fluctuations $\delta n/n_0 \gg 1 $ develop  even for mild wave amplitudes  $\delta_A \ll 1$
  \item Modulations develop faster for larger fluctuations  $   \delta_A$ (naturally) and larger $k_A$ (closer to the critical one); for  higher  magnetization $\sigma_A$ density   modulation develop slower.
     \item  Structure of density and \EM\  field modulations depend on magnetization $\sigma_A$, changing from 2-3 per period at mild $\sigma_A \sim 1$, to the high  $k/k_0 \sim 3 \times \sigma_A$ for $\sigma_A \gg 1$ (so that short wavelengths are quickly generated).  The final result  is generally independent of the value of the fluctuating field $   \delta_A$ (this is due to  the beat between the  forward and backward propagating waves in relativistically moving plasma in the \Alfven frame); for higher $\delta_A $ evolution  just proceeds faster.
  \item Spectra of density fluctuations and \EM\ fields are correlated, but  do not necessarily match. For example, at  a given time they may have different dominant modes; and/or overall power may peak at different times (density fluctuations peak earlier than that of the EM fields).
   \item There are  short periods of violent {\it longitudinal} (generation of  Langmuir waves) plasma relaxation. For CP wave in guide field,  even in pair-symmetric plasma, different charges  respond differently, leading to the generation of longitudinal plasma oscillations.
          \item The structure reaches a stable finite state, that depends on the  magnetization $\sigma_A$. In the final state, there is a number of new nonlinear waves per initial wavelength {\it and} density modulations (so, a new state is {\it not} a scaled version of the  initial state).
          \item CP and LP behavior is similar: the  final state preserves the initial polarization.
\end{itemize}

  \subsection{Basic run}
   
\begin{figure}
 \includegraphics[width=.3\linewidth]{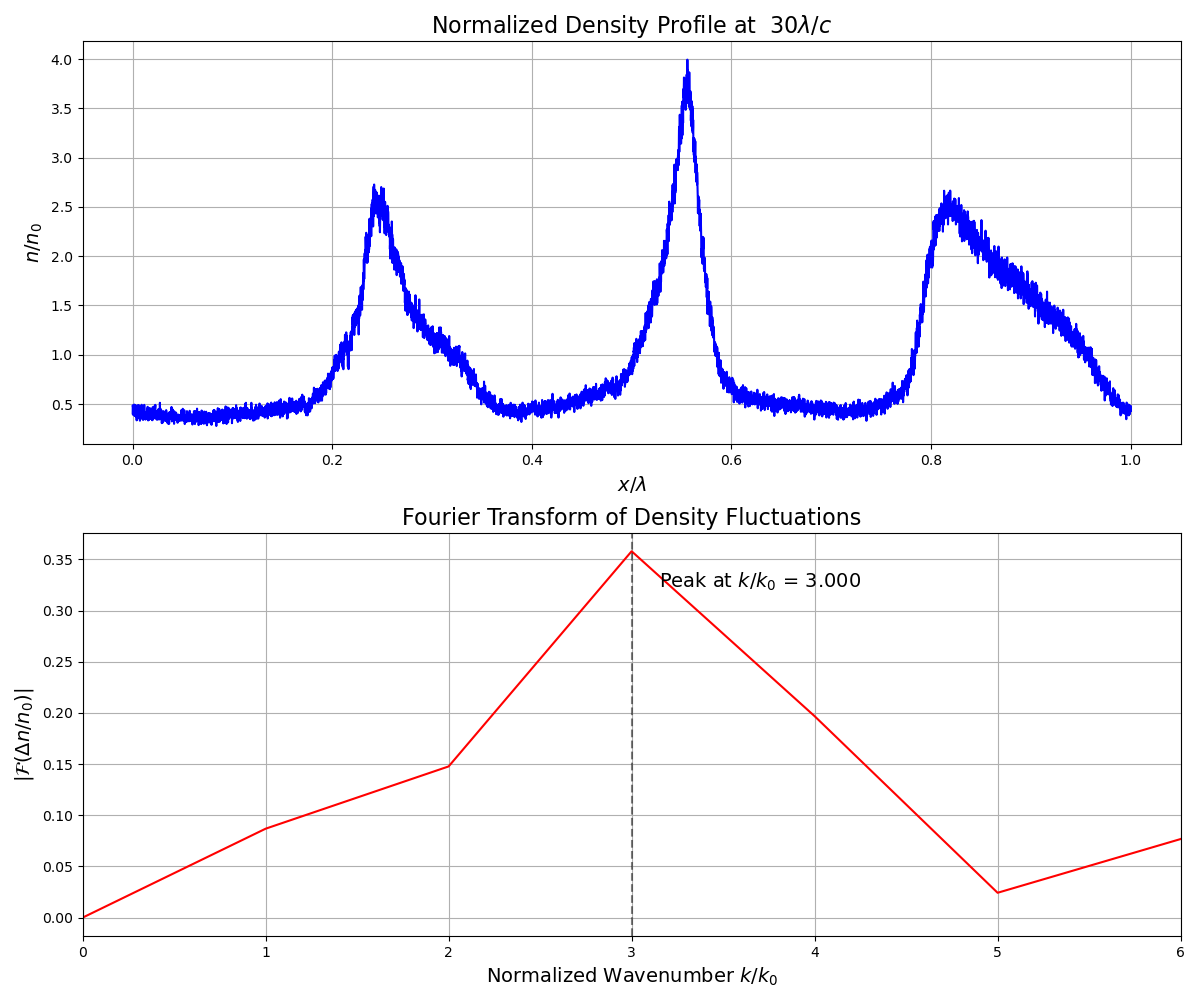}
 \includegraphics[width=.3\linewidth]{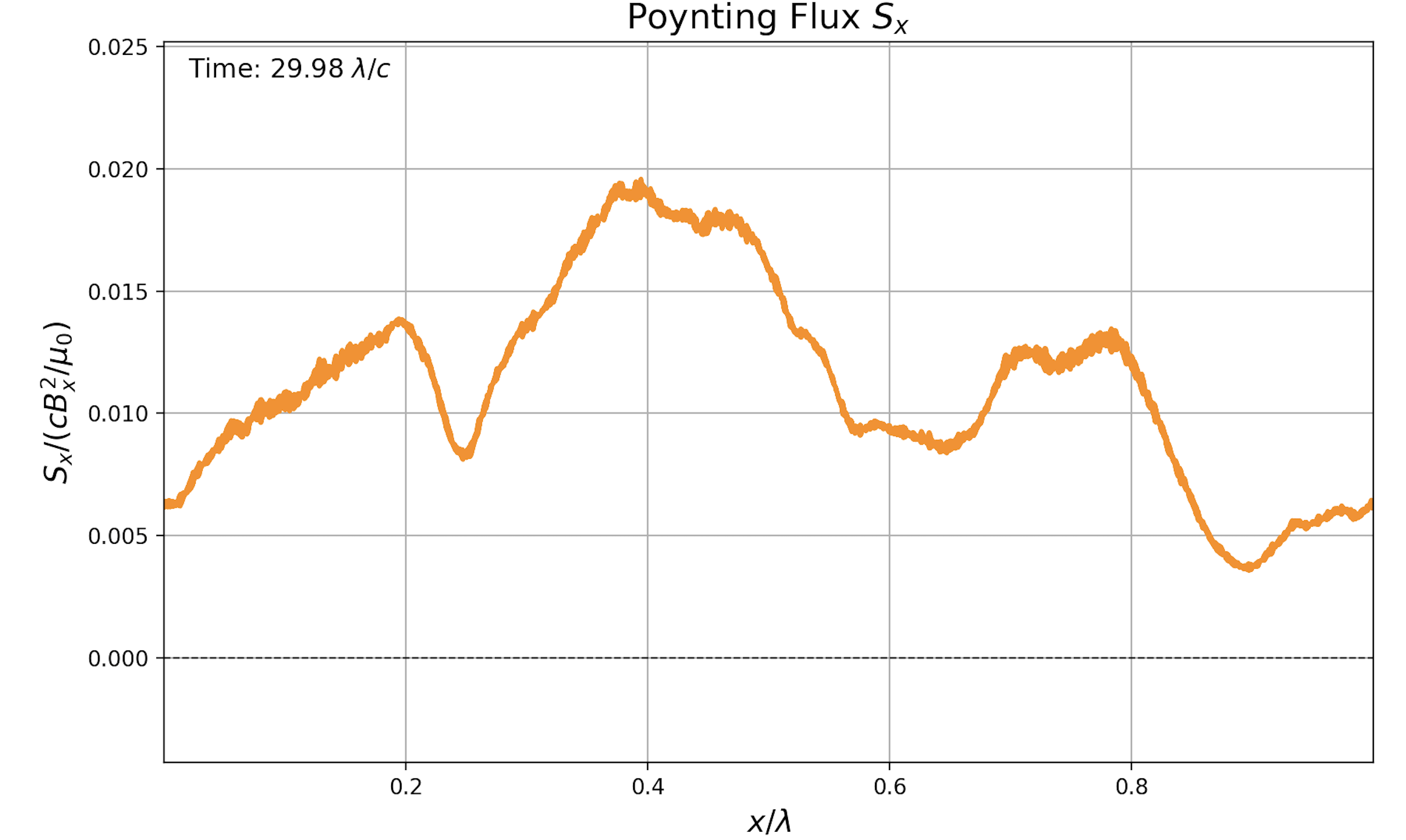}
  \includegraphics[width=.3\linewidth]{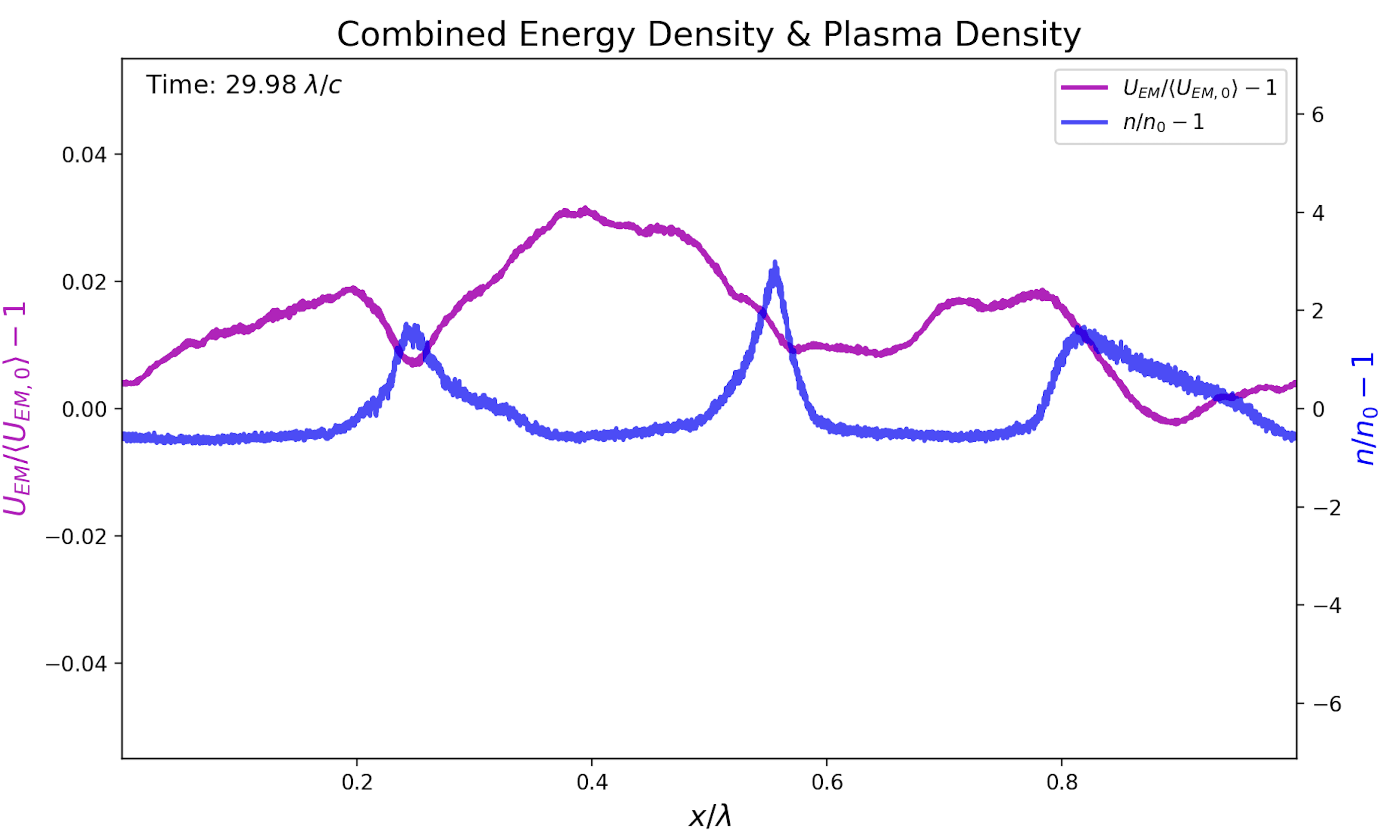}\\
   \includegraphics[width=.3\linewidth]{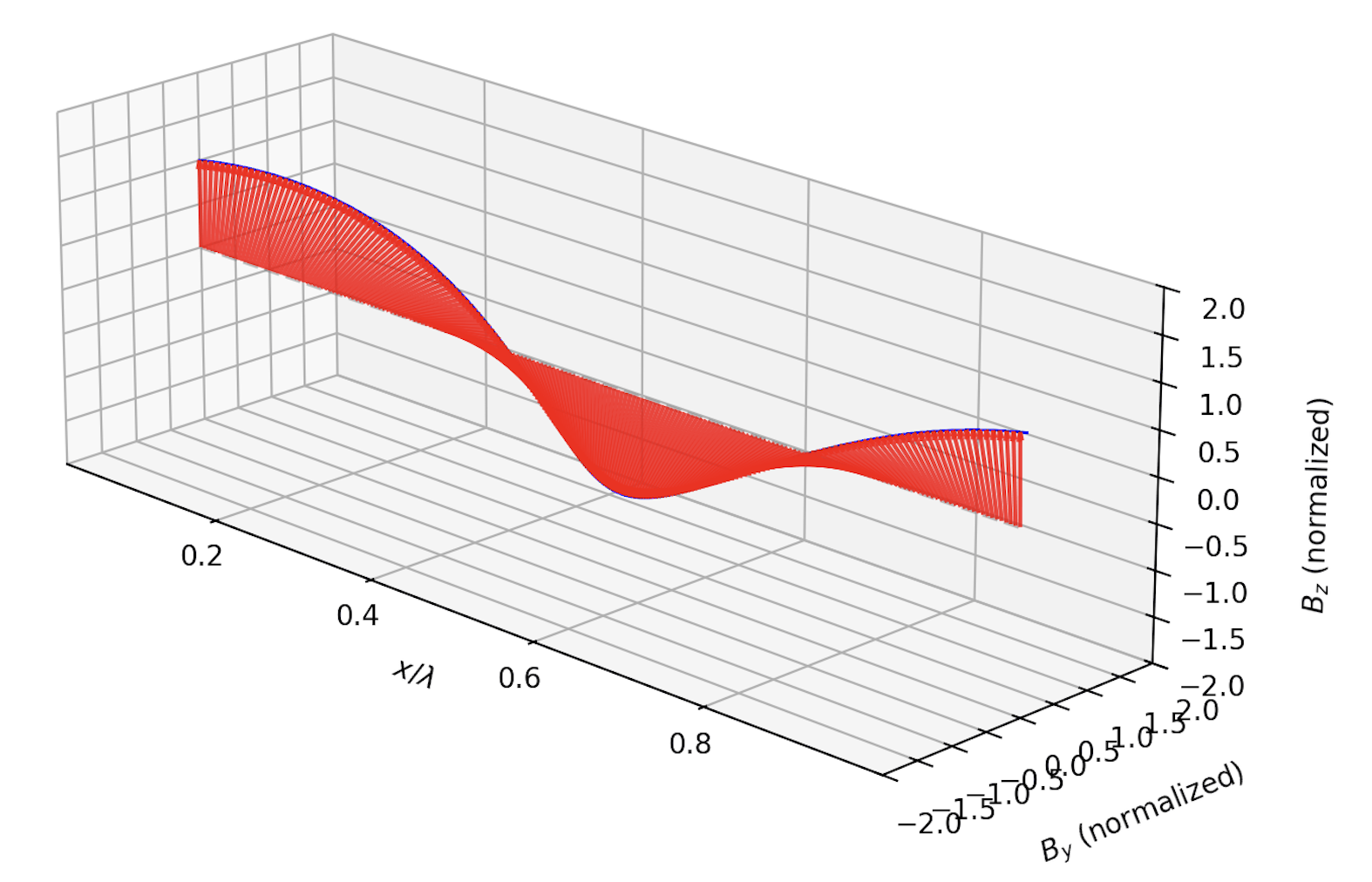}
     \includegraphics[width=.3\linewidth]{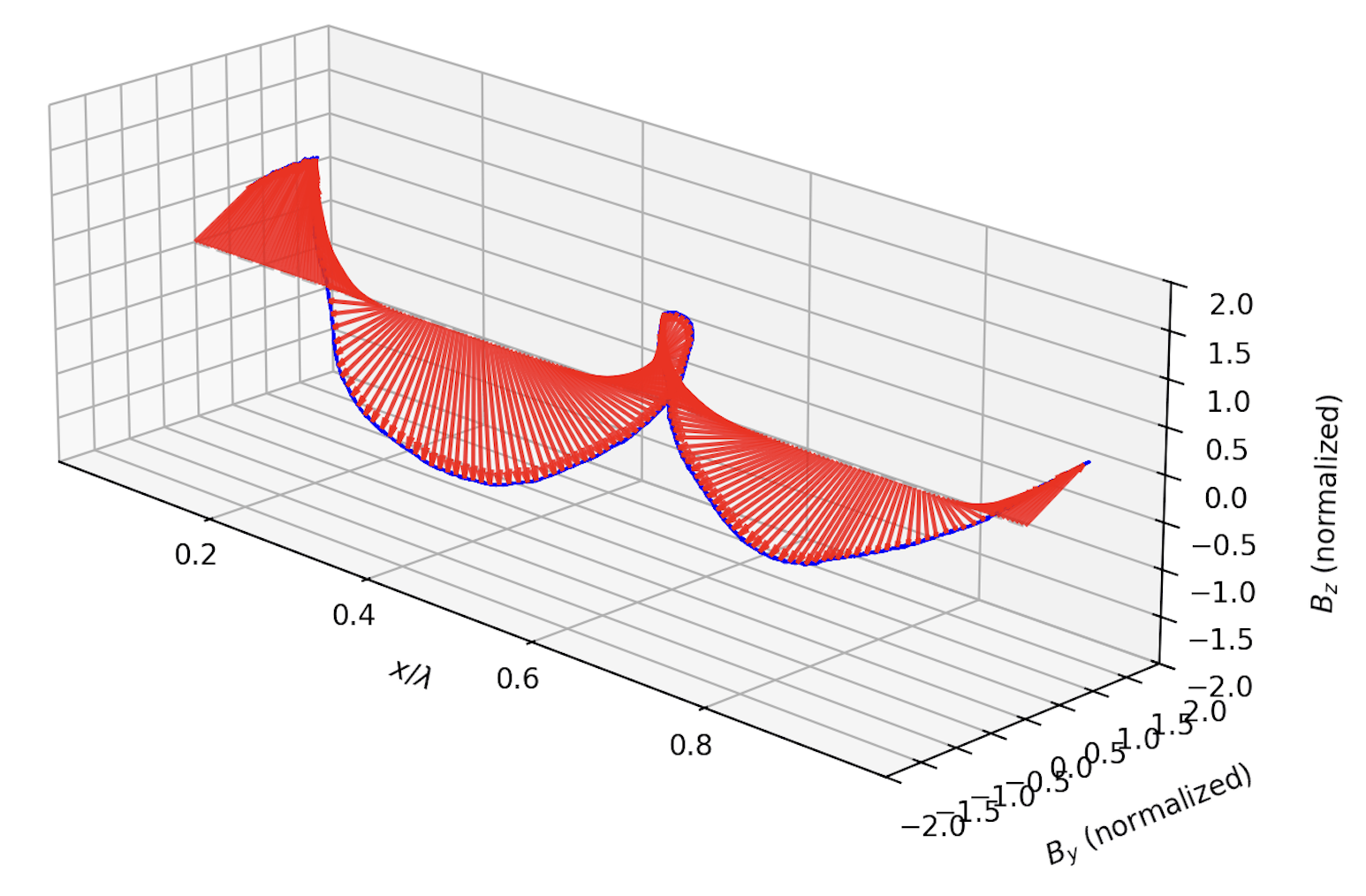}
       \includegraphics[width=.3\linewidth]{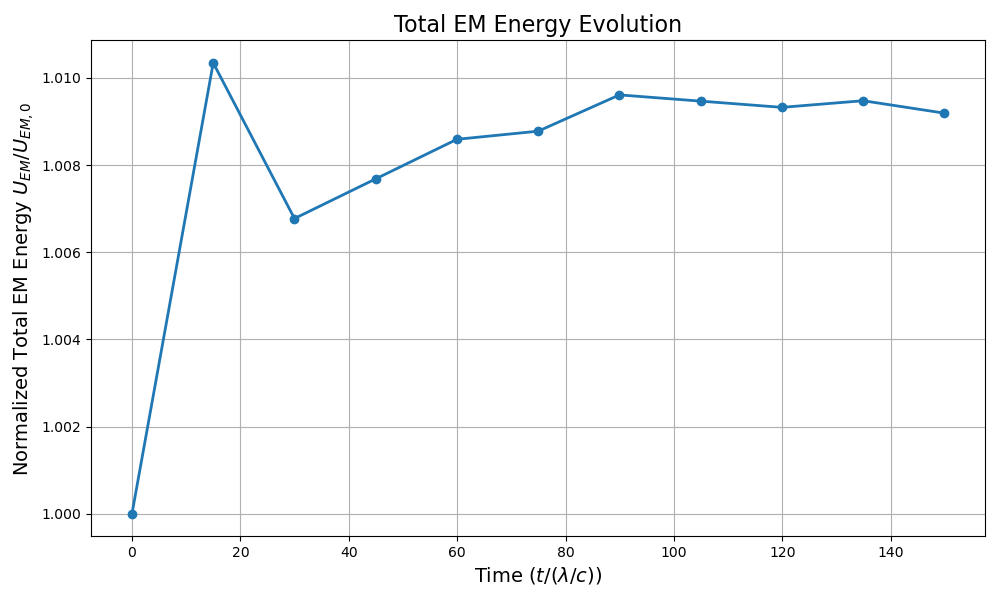}
\caption{ Single wavelength run, time $30 \lambda /c$, CP.  Plotted are normalized density profile and its Fourier transform,  Poynting flux (initially zero), combined plot of EM energy density and plasma density. Bottom row: initial and final polarization 3d rendering, and long-term evolution of \EM\ energy. }
\label{x1}
\end{figure}

   For a basic run, Fig. \ref{x1}, we use $\delta_A = 0.1 $,   ${\cal K}_A=0.1$ and  $\sigma_A =1$, box size equals one wavelength. Corresponding \Alfven momentum is $p_A = 0.795$.
   
We observe that a wave separates (quickly, on time scale of $\sim$ few $c/\lambda$),  into three sub-waves, exactly at  wavelength of $1/3$. There are large density fluctuations between the sub-waves',  peaks of densities correspond to minima of EM energy - this is due to the ponderomotive force pushing plasma particles  out of high \EM\ density energy regions   \citep[similar to the effects observed in Refs. ][]{2025arXiv250920594L,2025arXiv250917245L}. The wave, initially at rest  (initially zero Poynting flux) is moving to the right, that is, its phase velocity increased. New structure keeps the  circular polarization. 

The final structure is persistent (not breaking-up further). It is not just a reduced version of the initial state: there are large density barriers holding/trapping  the localized waves.The plot of total EM energy (longer integration times) show that a system reached a steady state, Fig. \ref{x1}.

\begin{figure}
     \includegraphics[width=.45\linewidth]{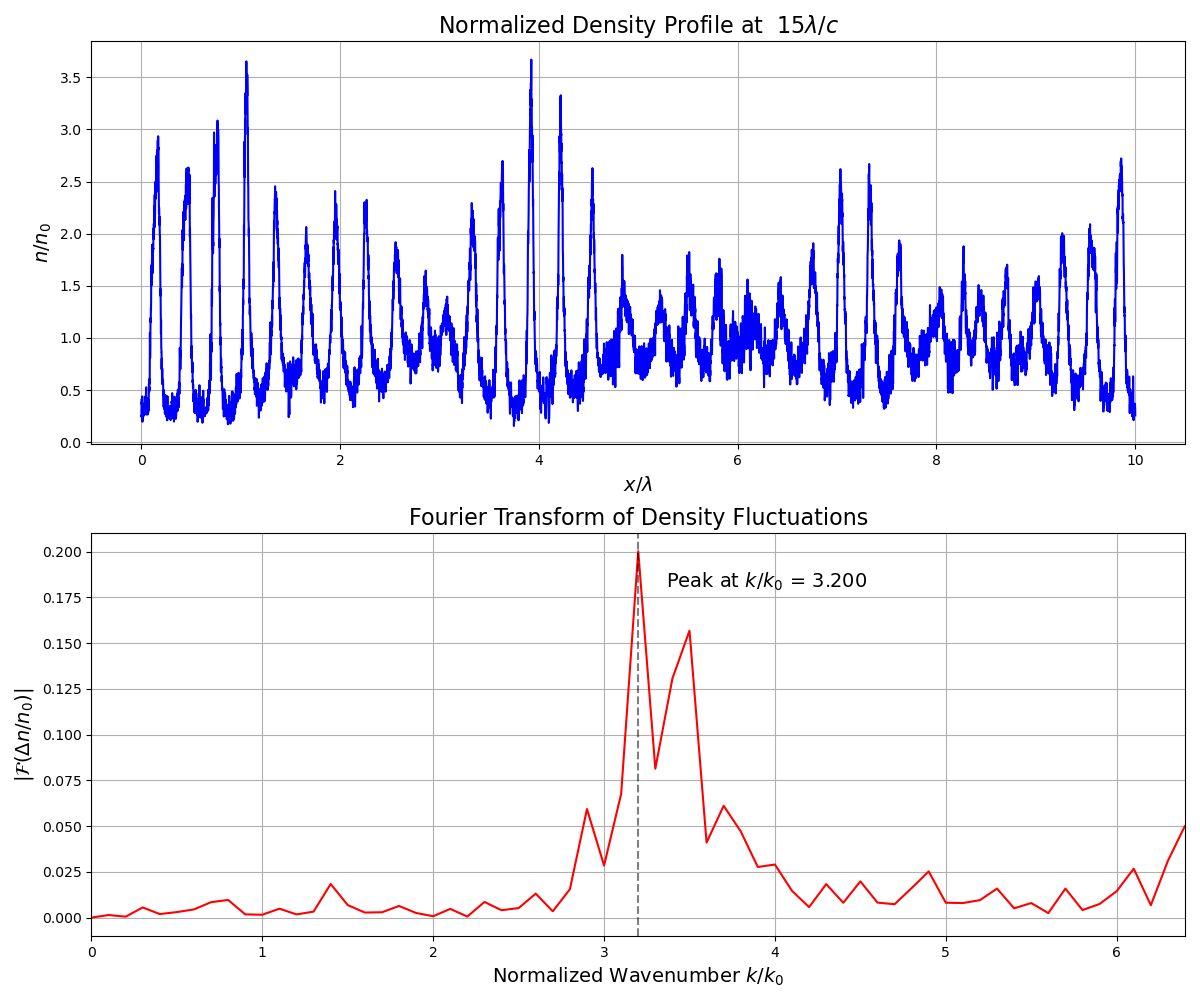}
      \includegraphics[width=.45\linewidth]{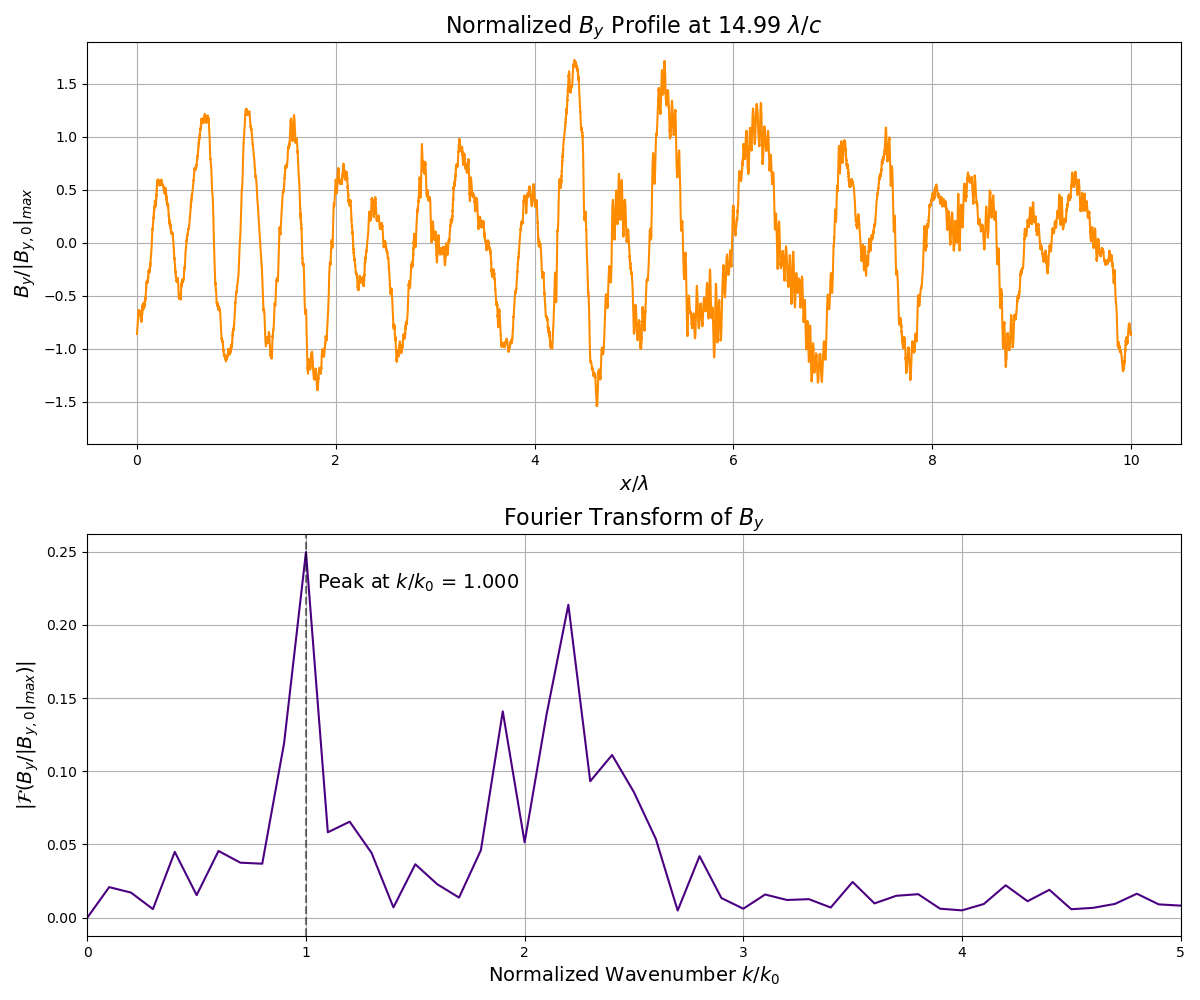}
        \includegraphics[width=.45\linewidth]{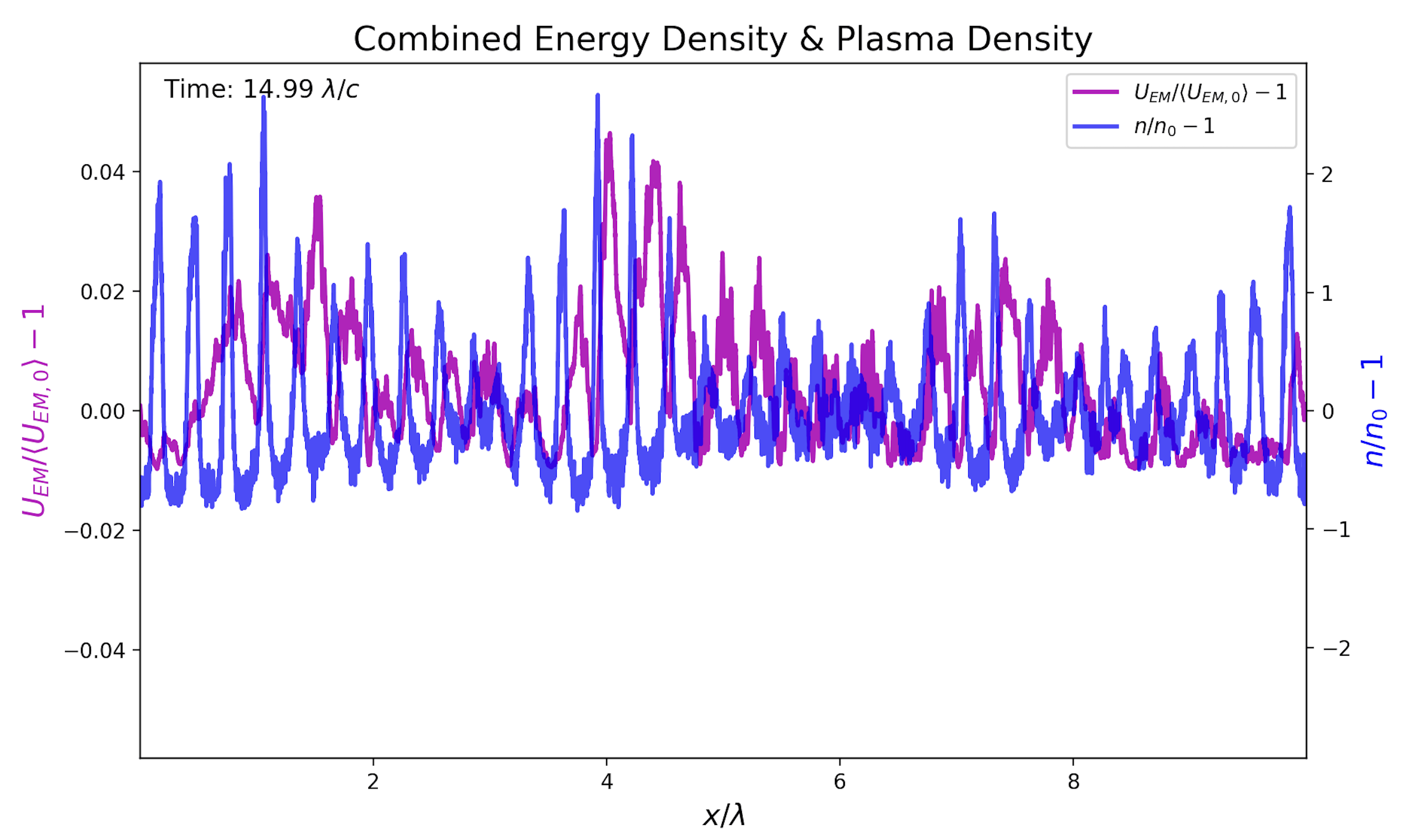}
          \includegraphics[width=.45\linewidth]{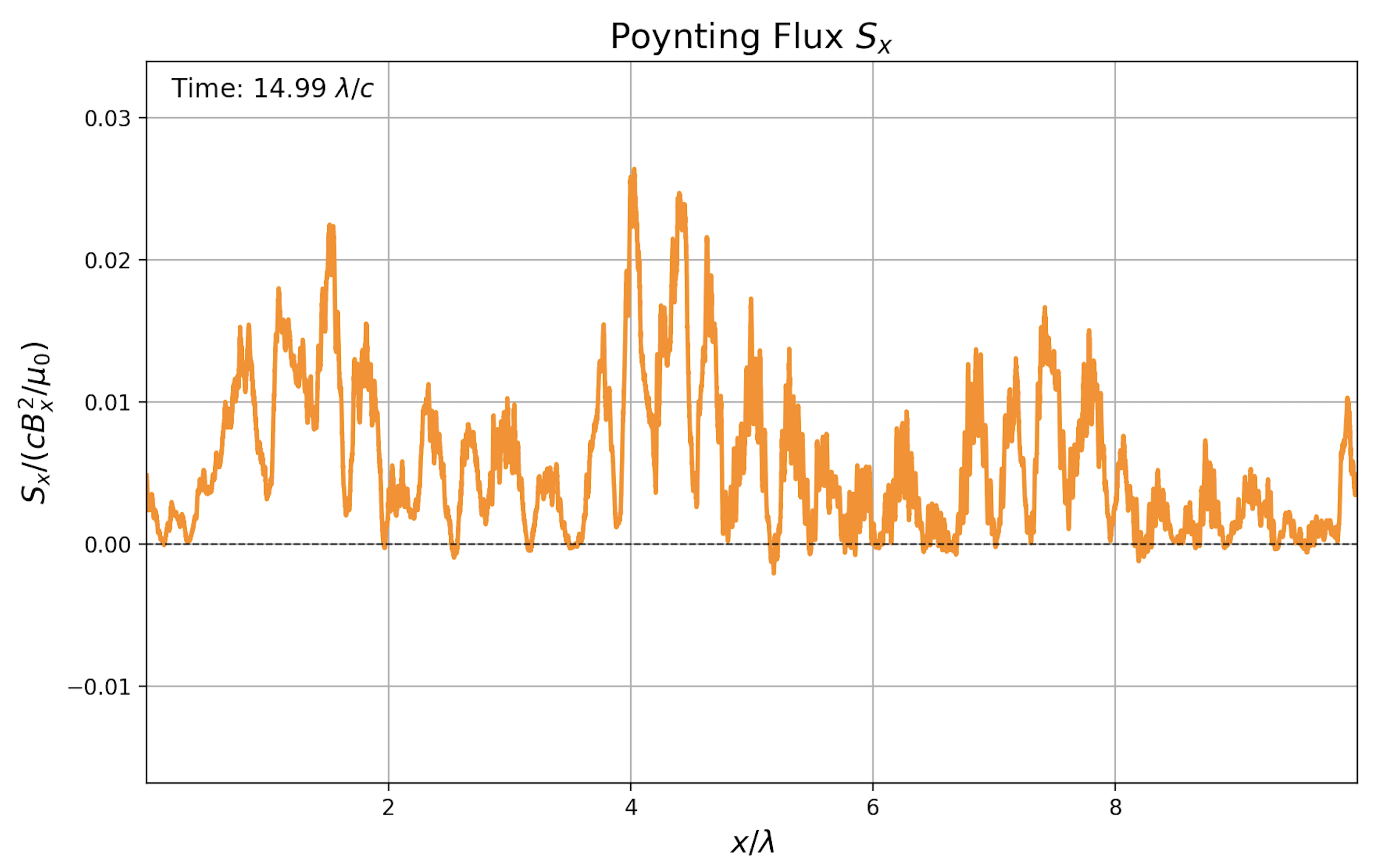}
\caption{ Simulations of a longer wave train, $10 \lambda$. Density and its Fourier transform, magnetic field $B_y$ and its Fourier transform, combined plot of density and \EM\ energy density (peaks of \EM\ density generally correspond to troughs of plasma density), and Poynting flux. Positive pointing flux indicates that modulated waves propagate faster than the driver (in the initial configuration the Poynting flux is zero).  Similar to the single wavelength simulations, the dominant  density peak at $\approx 3 k_0$.}
\label{long}
\end{figure}

\begin{figure}
 \includegraphics[width=.45\linewidth]{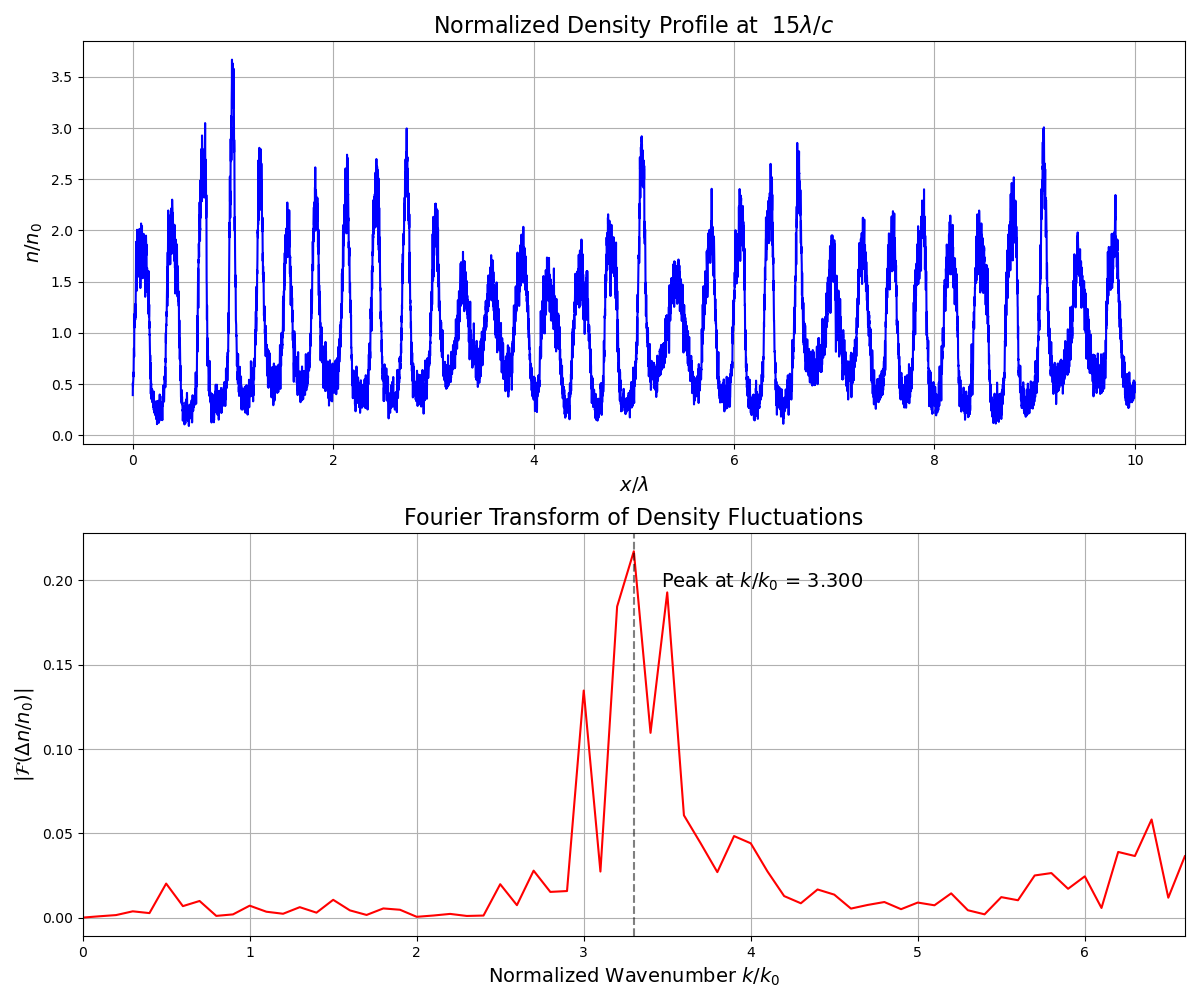}
  \includegraphics[width=.45\linewidth]{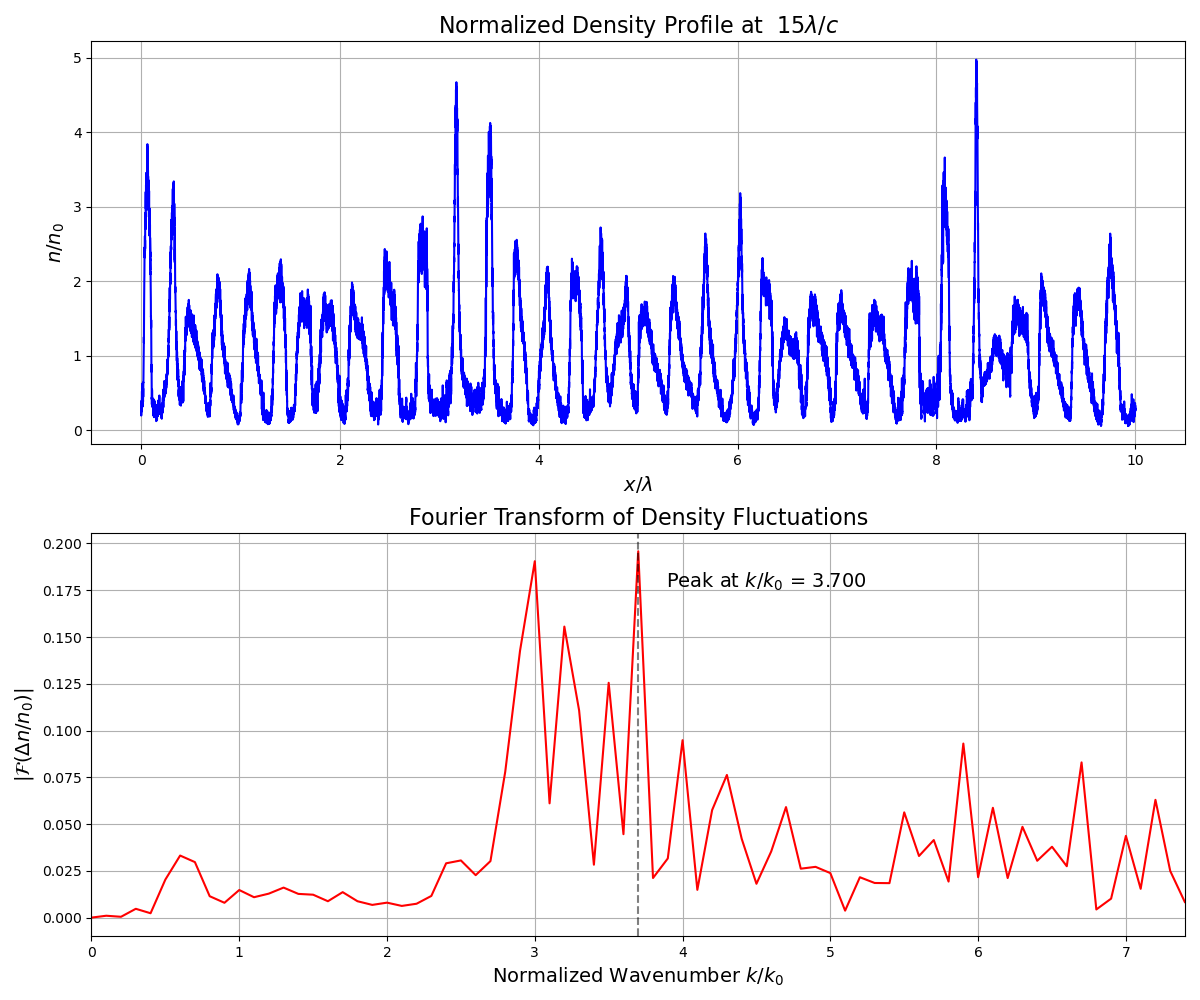}
 \caption{Comparison of resolution for larger $10 \lambda$ simulations, Basic Run:  left panels  $n_x=100 \times 30$, $n_p=30$, right panels $n_x=100 \times 300$, $n_p=100$. For higher resolution there is somewhat more power at higher $k$, but results are generally consistent (eg, both show a peak at exactly $3 k_0$. Results are generally consistent with the  range of   $ \lambda=1$ simulations. }
\label{compare-resolution}
\end{figure}

Large density fluctuations are the key, specific component  of pair plasma \citep{2025arXiv250920594L}. 
Generation of large density fluctuations is not specific to the one-wavelength runs, Figs  \ref{long}-\ref{compare-resolution}. 

\begin{figure}
     \includegraphics[width=.3\linewidth]{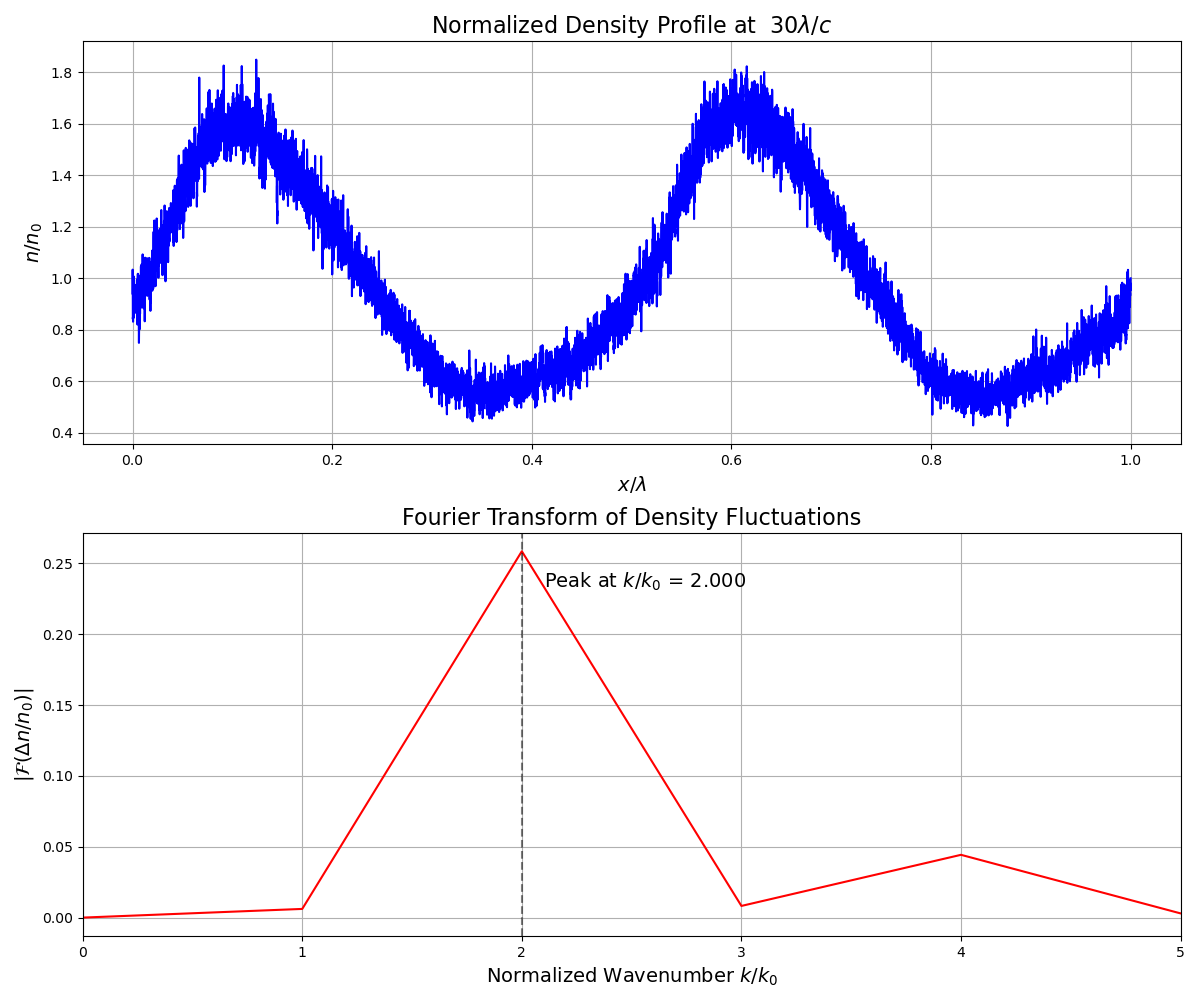}
 \includegraphics[width=.3\linewidth]{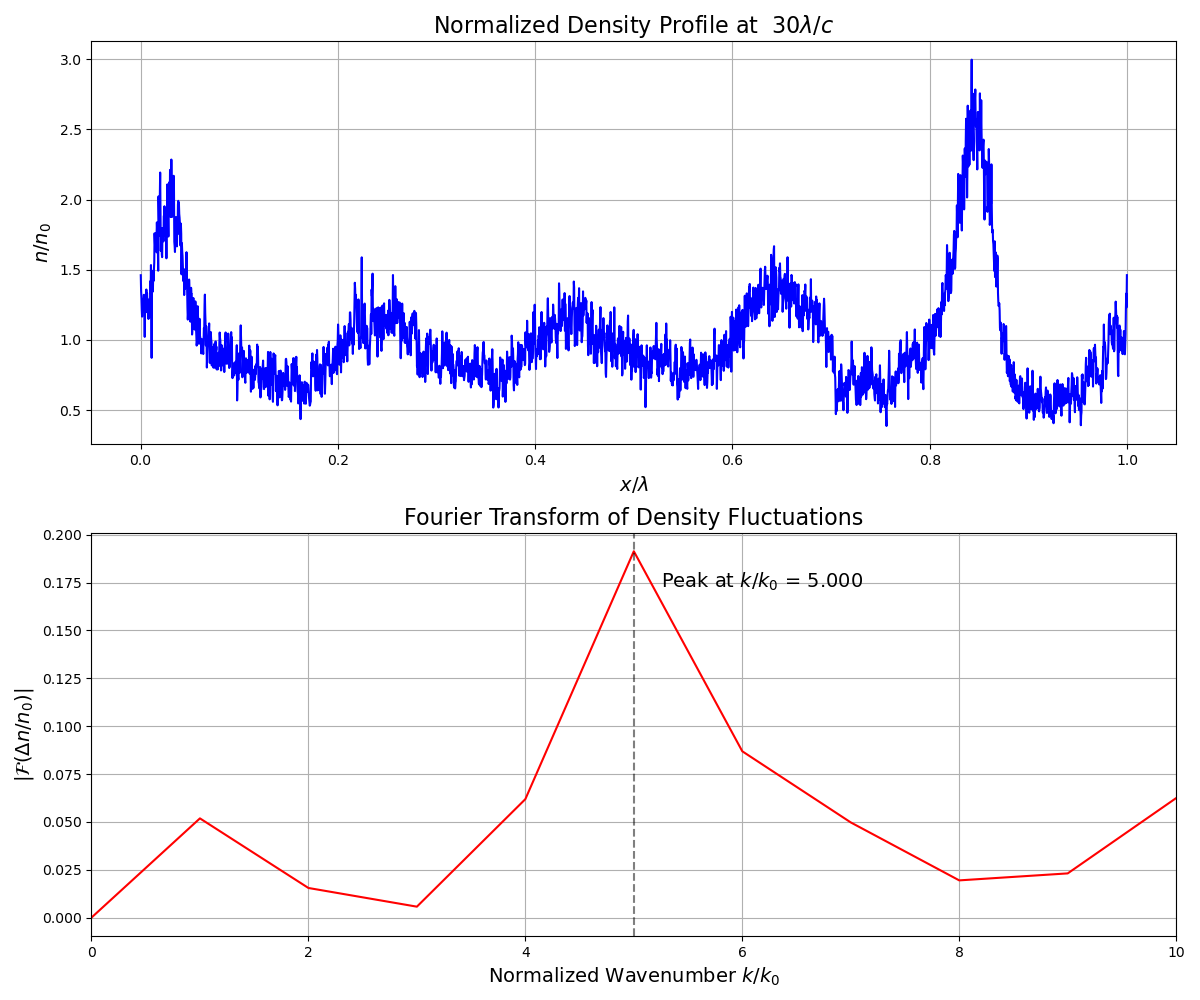}
  \includegraphics[width=.3\linewidth]{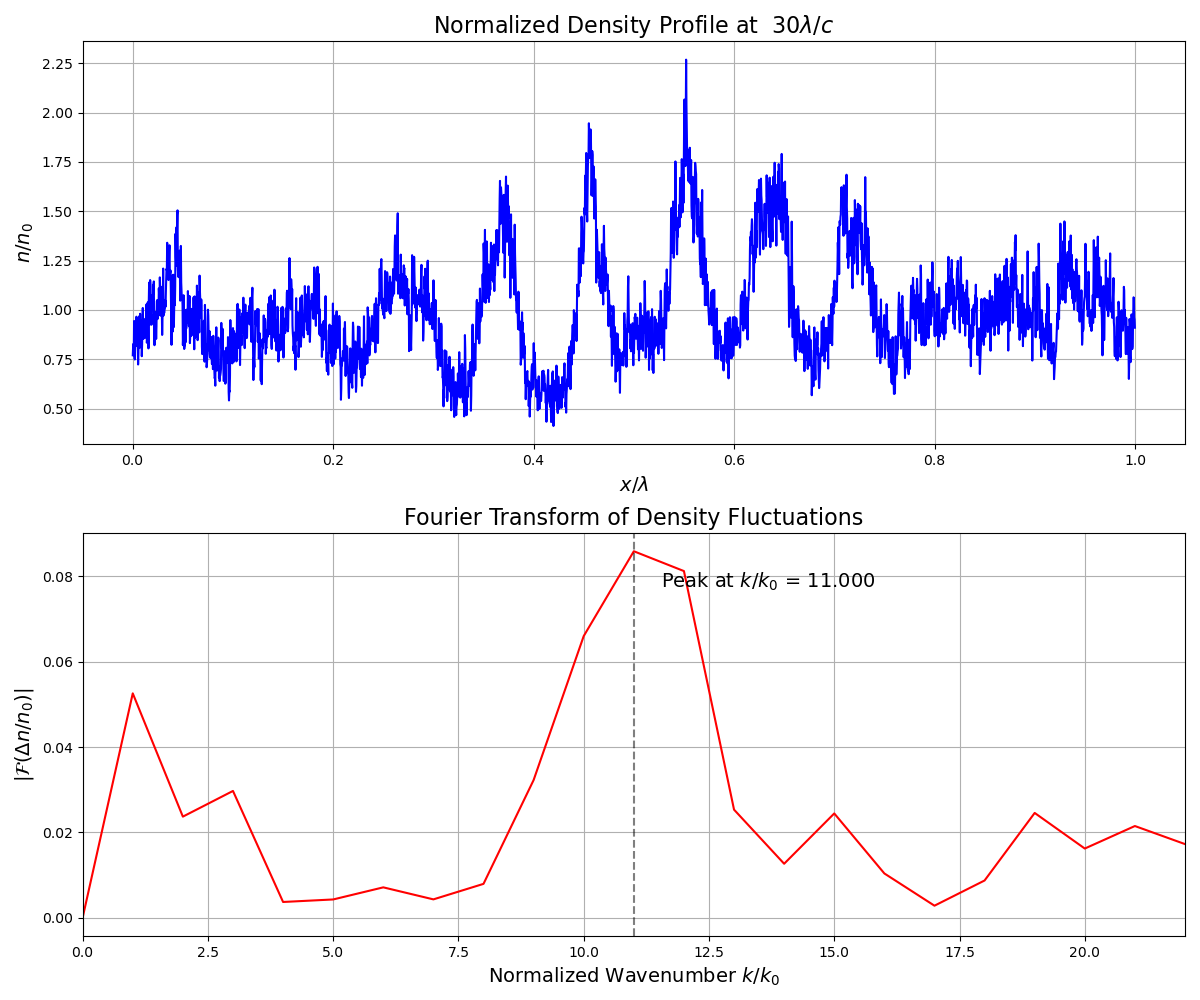}
\caption{ Sigma-dependance of modulation:     $\sigma_A =0.1 $  (left)  wave modulates into 2, $\sigma_A =2$  (center)  wave modulates into 5,   $\sigma_A =5$  (right)  wave modulates into 11.}
\label{x1-sigma2}
\end{figure}

The  number of resulting subways is $\sigma$-dependent. In  short one-wavelength runs,  it  is always  an integer number, Fig. \ref{x1-sigma2}. For higher sigma, the break up occurs faster, and into more numerous sub-waves.

At low $\sigma$ runs, {\it e.g},  at $\sigma_A =0.5$ we observe first formation of triple density structure, that gets destroyed, and the system relaxes back to the equilibrium.  Our exploration of whether this is a numerical effect due to finite resolution did not produce solid conclusion. 

Generally, formation of periodic density structures in lower-sigma runs does not lead to considerable wave modulation: long-living double-density walls slightly modify the wave:  the wave's wavelength and amplitude remain approximately the same  (but the wave starts moving slowly  in the initial \Alfven frame due to a somewhat  different dispersion, )

\begin{figure}
     \includegraphics[width=.45\linewidth]{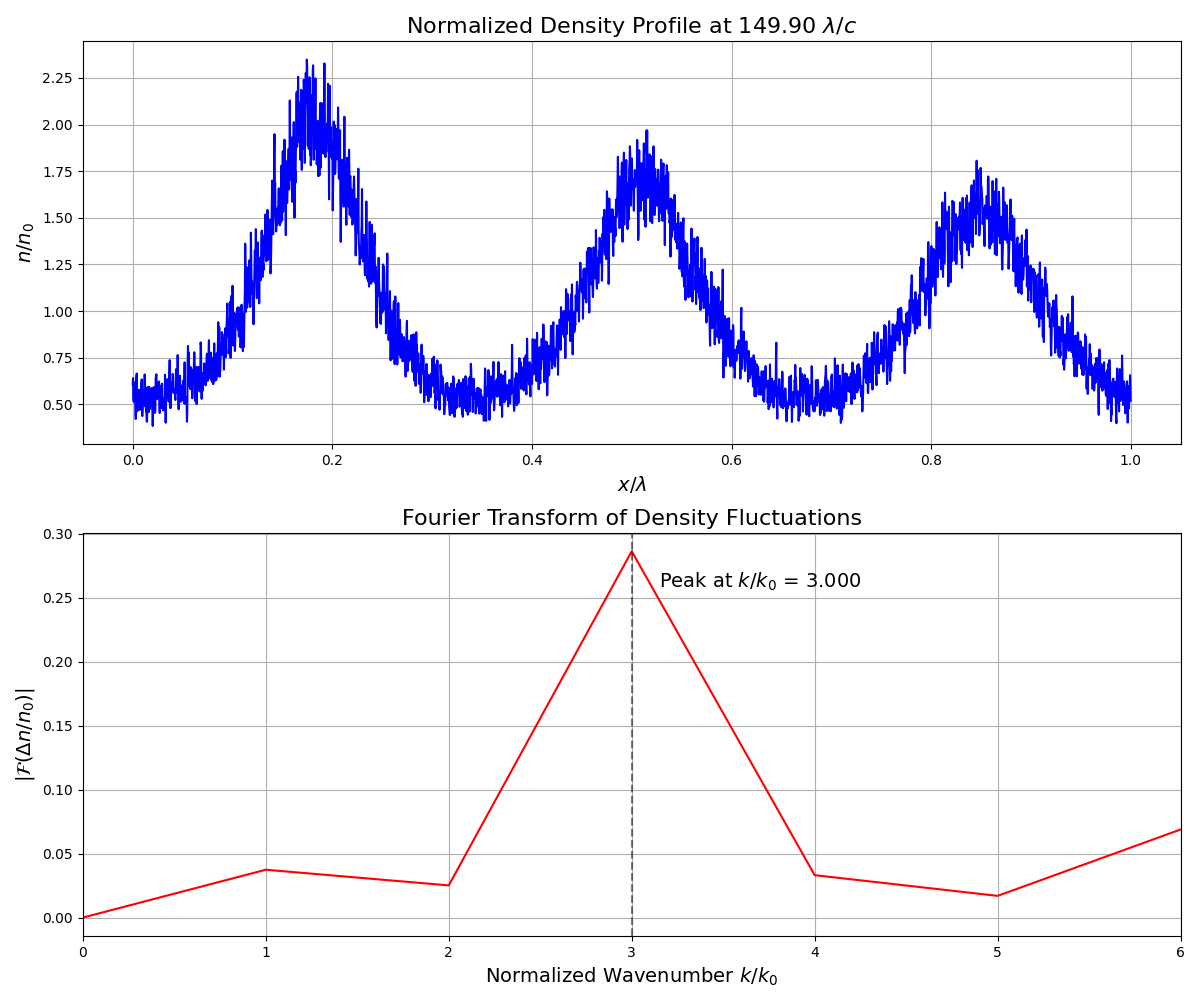}
      \includegraphics[width=.45\linewidth]{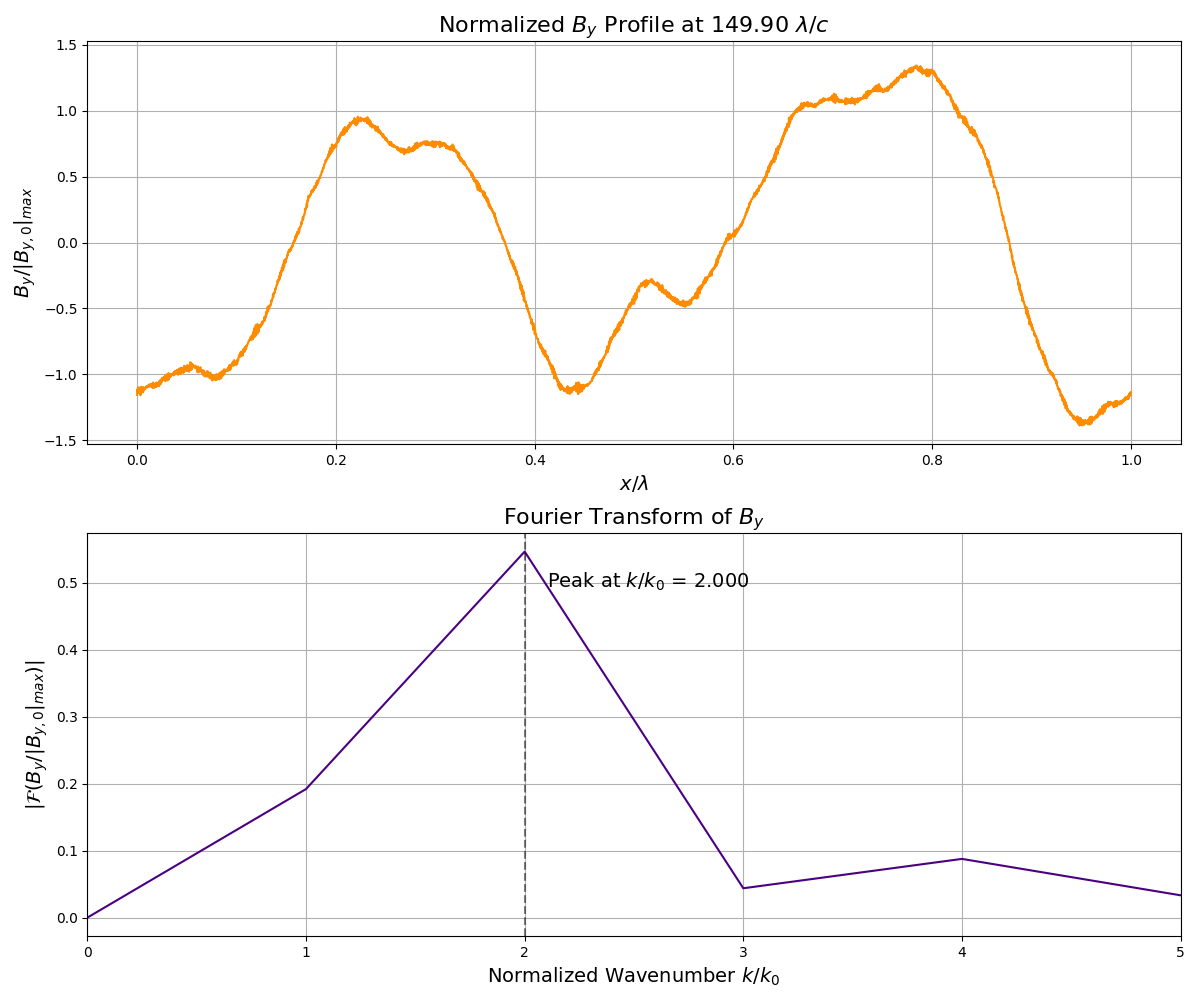}
\caption{ Modulations of density and EM properties may occur at different wavelength. In the example, density is modulated at $3 k_0$ while the wavelength is $2 k_0$. (newly created pattern is not stationary in the initial frame).}
\label{dens-EM}
\end{figure}

In the higher-sigma runs, density modulation does lead to EM modulation, but not necessarily at the same wavelength, Fig. \ref{dens-EM}.

Periodic boundary conditions establish resonator-type behavior for the perturbations. But in \Alfven frame the  backward and forward propagating waves have different velocities (since one is propagating with the flow, and another  against it). For small magnetization, when the \Alfven velocity is slow the resonance occurs at the Bragg's condition of $2 k_0$.

  \subsection{LP \& thermal effects}
\begin{figure}
 \centering
     \includegraphics[width=.45\linewidth]{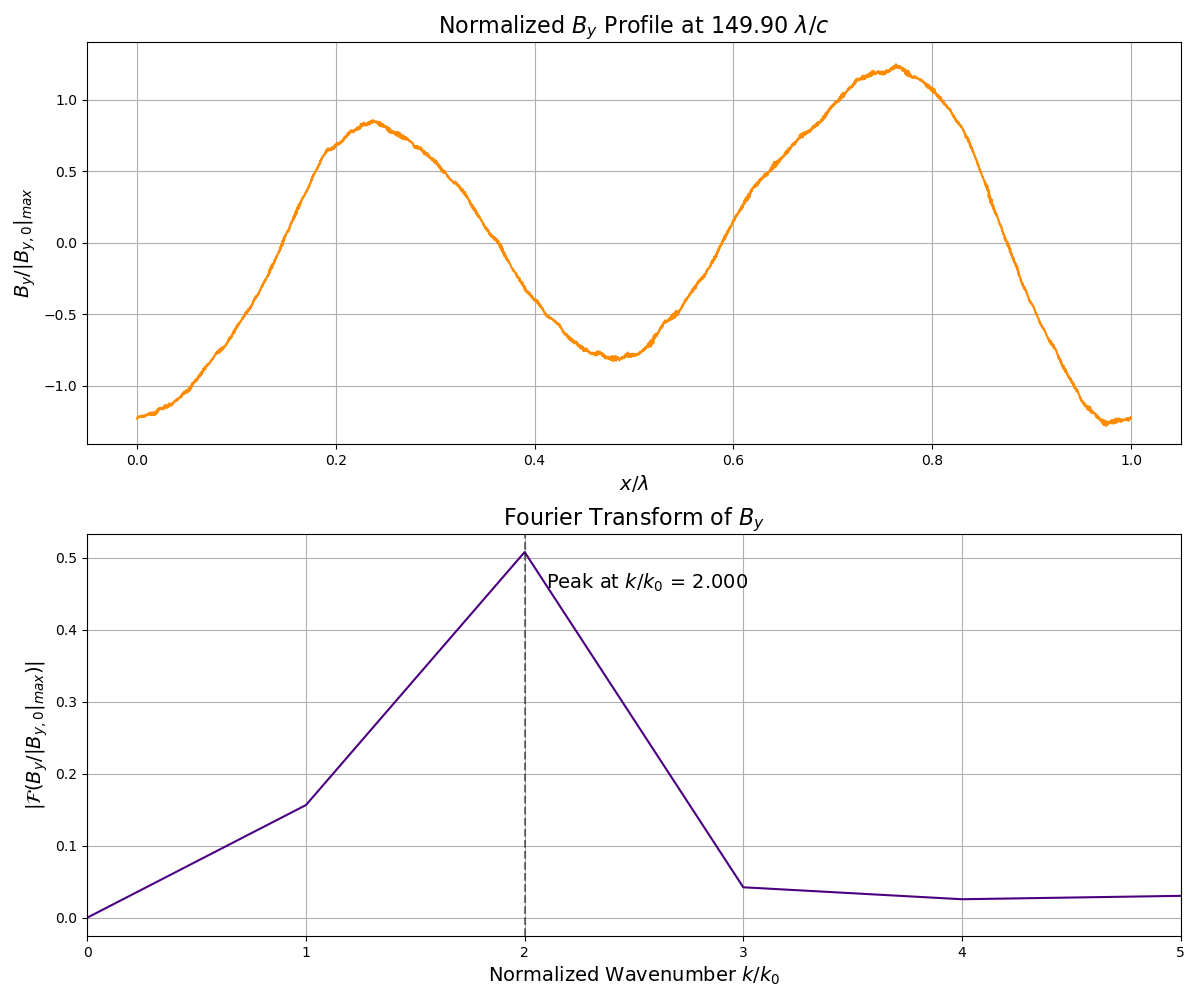}
     \vline
        \includegraphics[width=.45\linewidth]{./figures/lp/By_fft_0010-lp.png}
           \includegraphics[width=.45\linewidth]{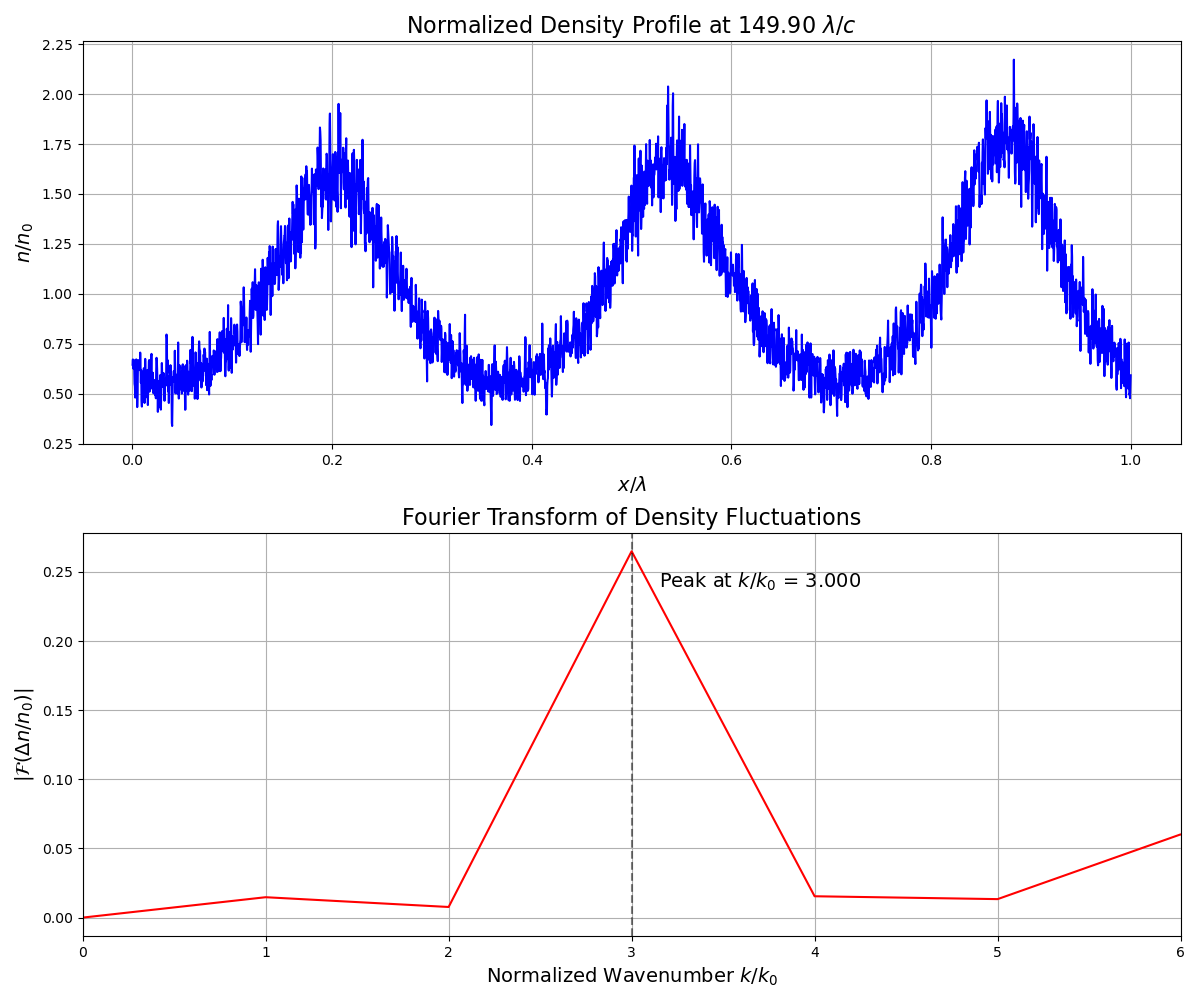}
           \vline
                  \includegraphics[width=.45\linewidth]{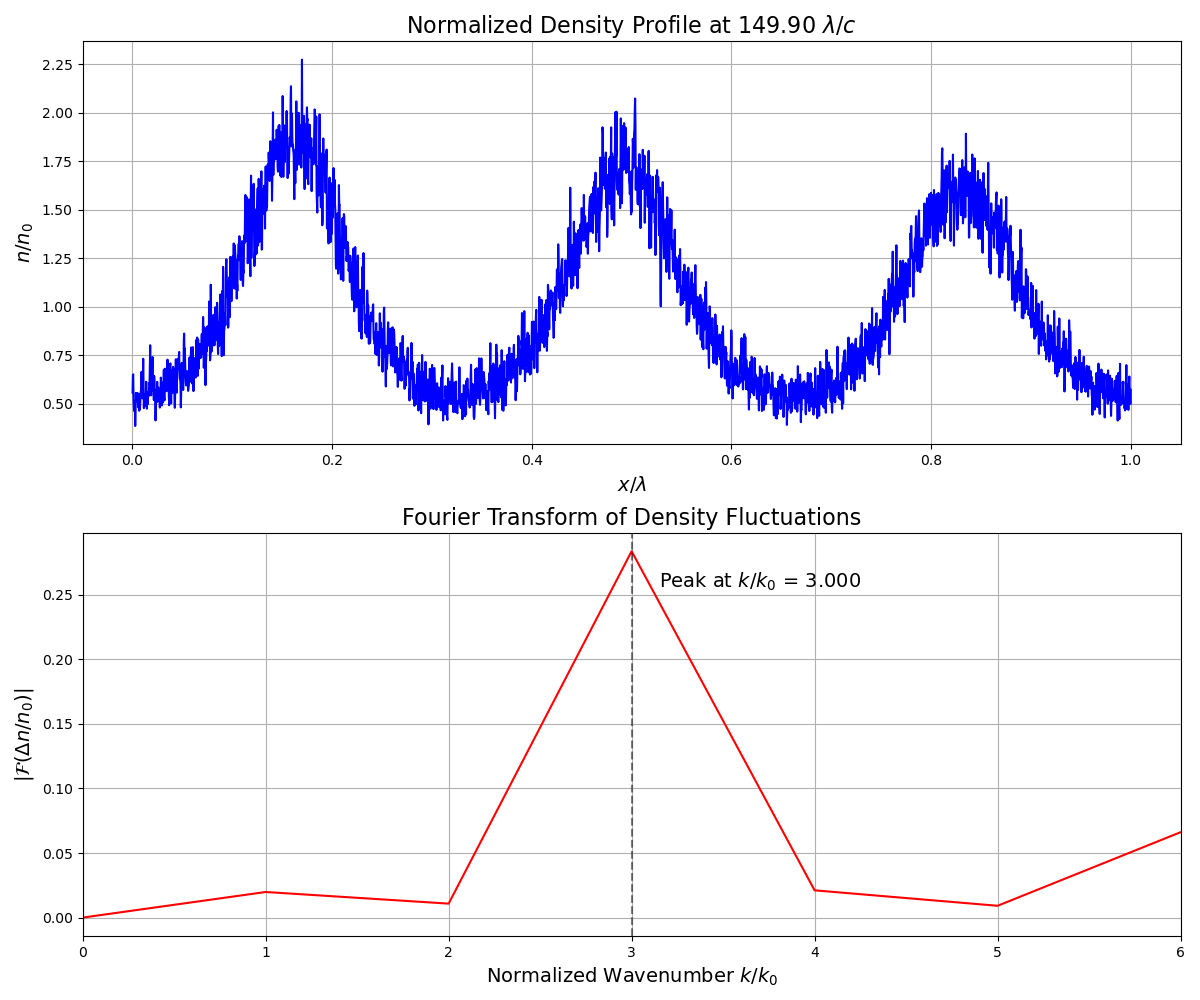}
                        \includegraphics[width=.45\linewidth]{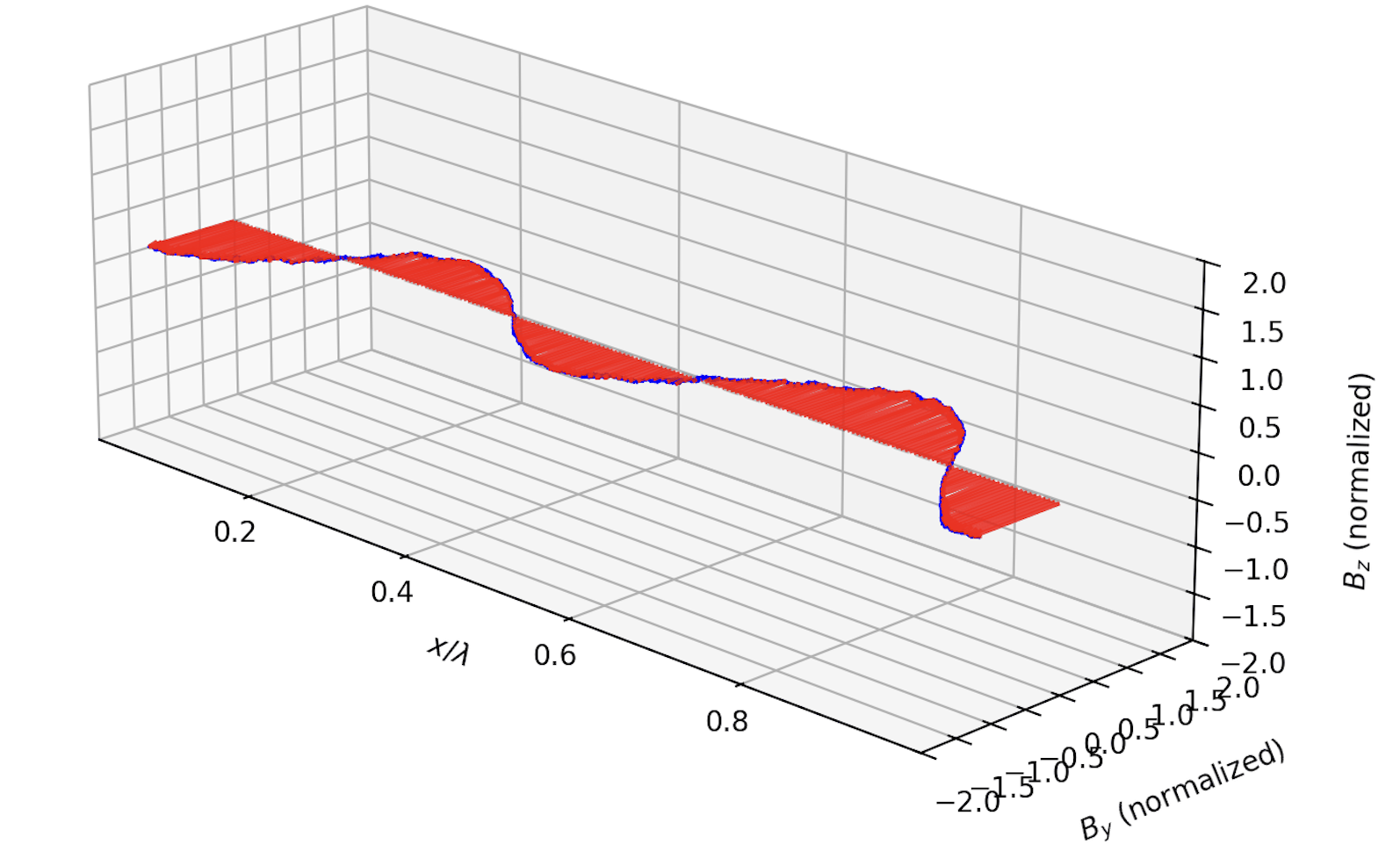}
                          \vline
 \includegraphics[width=.45\linewidth]{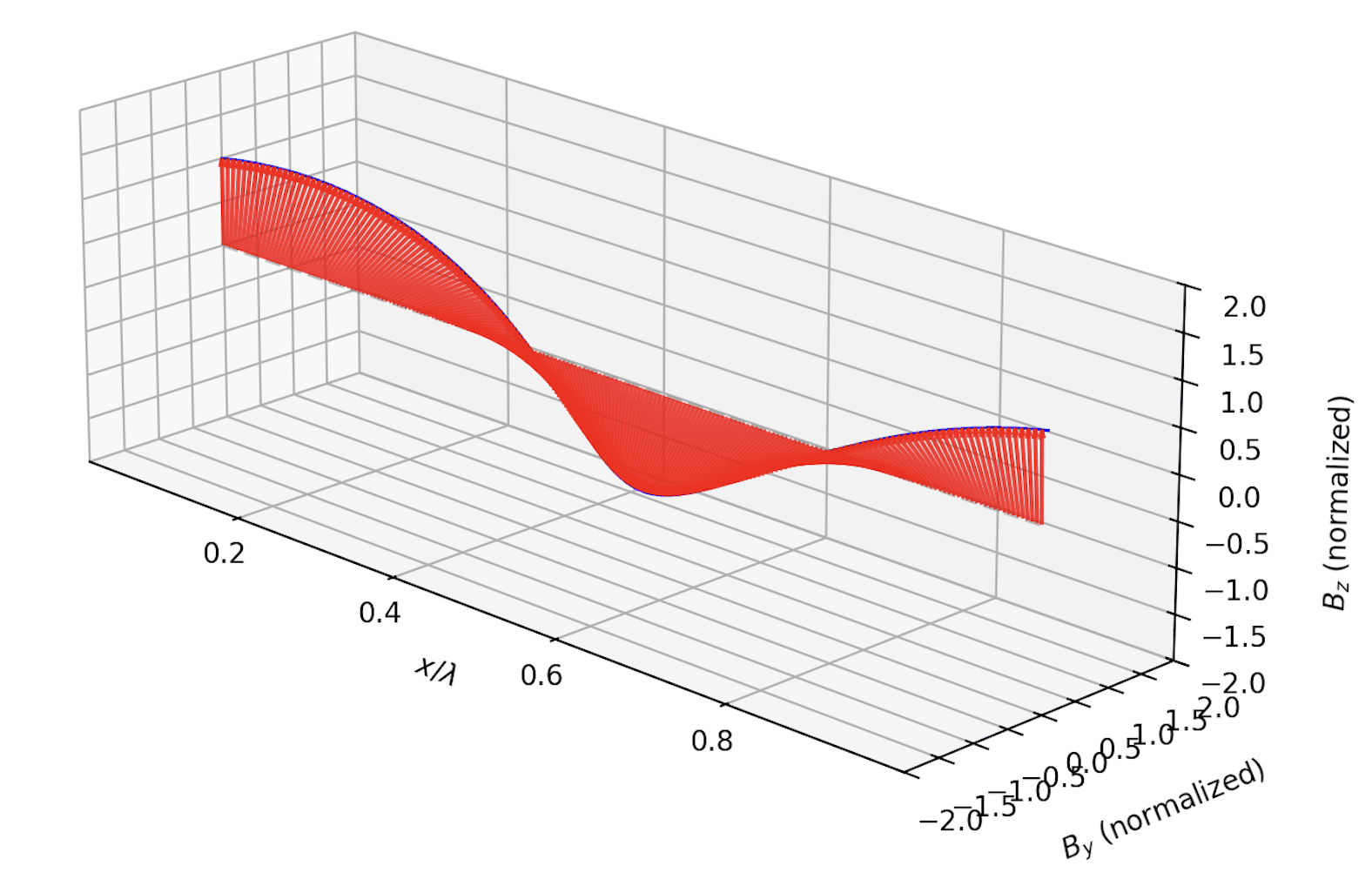}
      \caption{ Comparing LP (left column) and CP (right column) }
\label{lp}
\end{figure}

  For linear polarization (LP),  
analytically, only small amplitude waves   $   \delta_A \ll 1$ can be initialized self-consistently.   We choose (In the wave's frame \Ef\ is zero, all quantities are time-independent. ) 
   We chose 
   \ba && 
   {\bf e}_B = \left\{0,\sin \left(k_A x\right),0\right\}
   \nn &&
   {\bf B} =(  {\bf e}_x +  {\bf e}_B    \delta_A) B_0 
   \nn &&
   v_\pm = \left\{ \frac{  p_{A}}{ \gamma_A}  ,
   \sigma_A \frac{   \delta_A}{2 p_A}  \sin \left(k_A x\right),
   \pm \sigma_A \frac{   \delta_A}{2 }  \frac{k_A}{\om_B} \sin \left(k_A x\right) \right\}
   \ea
There is no charge separation. The bulk Alfven momentum $p_A$ is determined from the same linearized  equation as for the CP, Eq. (\ref{disp1}). Linear and circular polarizations behave similarly, Fig. \ref{lp}.   This has a clear explanation: an  LP wave is a sum of two CP waves, each producing density modulations on their own. The long-term evolution, shown in Fig. \ref{lp} is the similar between LP and CP.  

\begin{figure}
 \centering
     \includegraphics[width=.49\linewidth]{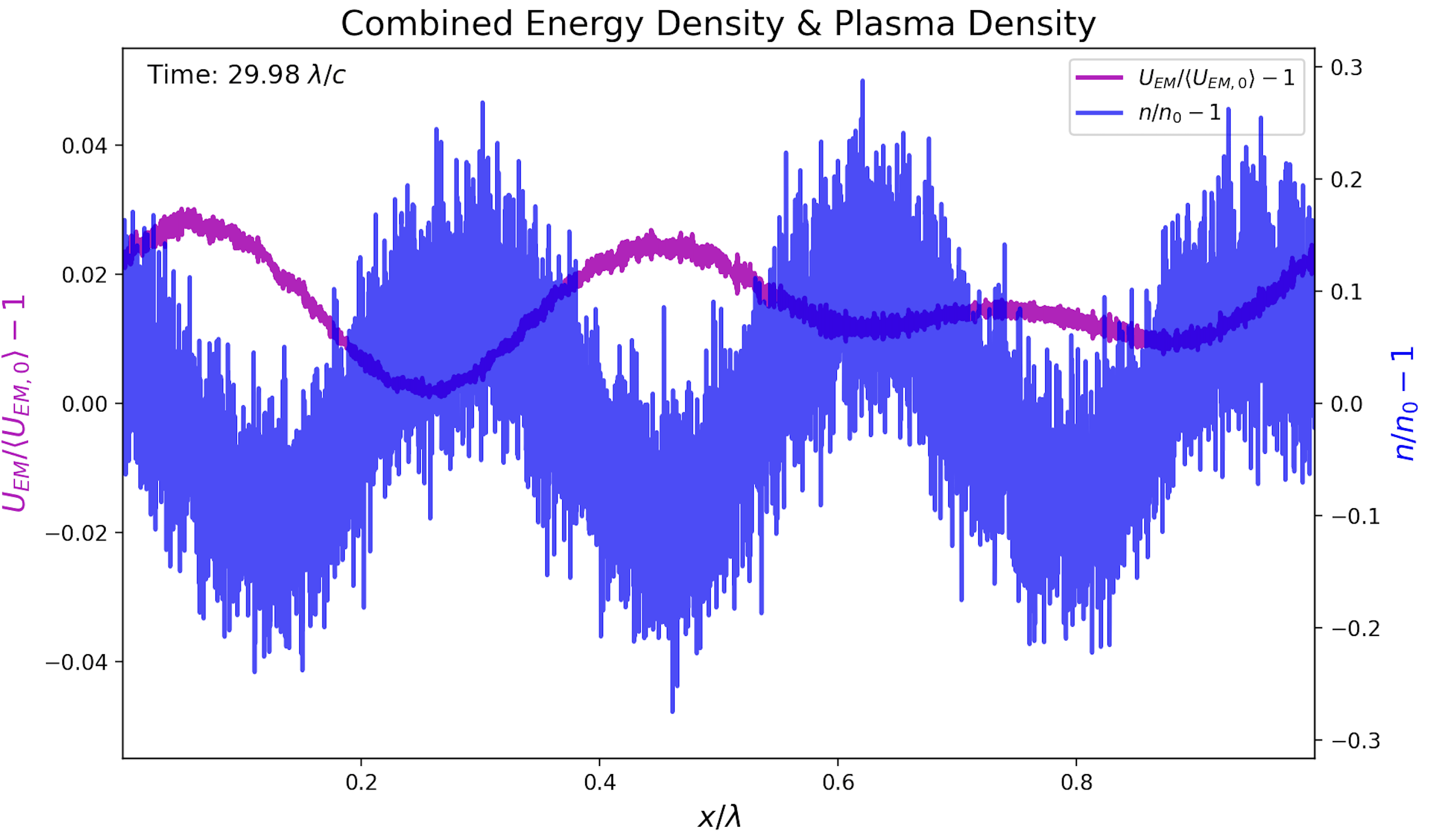}
        \includegraphics[width=.49\linewidth]{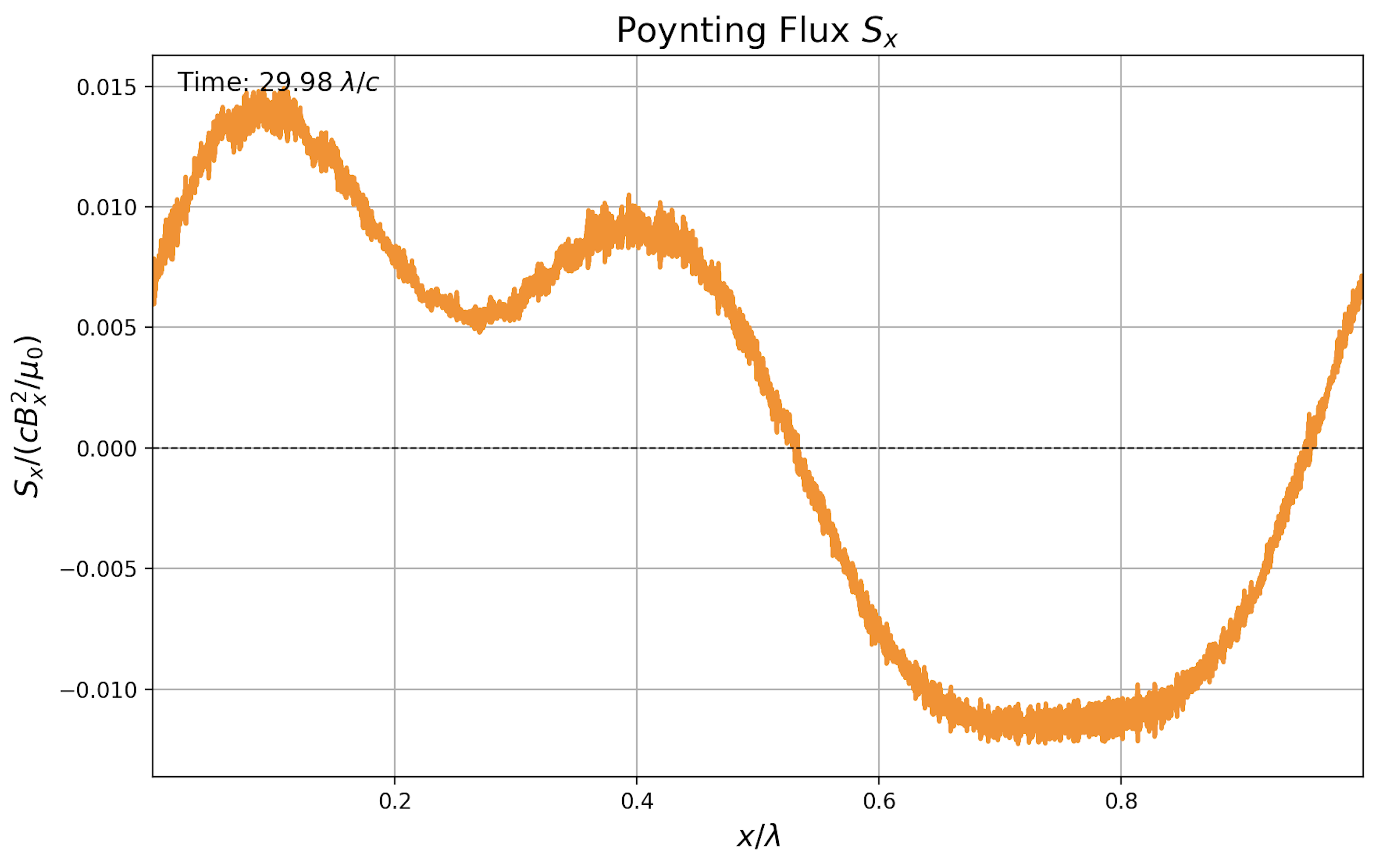}
       \caption{Suppression of modulational instability by thermal effects. For this run $T/(m_e c^2) = \delta_A^2$. Though modulation can still be seen, the absolute amplitudes of both \EM\ and density fluctuations are }
\label{hot}
\end{figure}

We also verified that, as  expected, thermal motion suppresses the modulation, Fig. \ref{hot} 

\subsection{Transient Langmuir oscillations}

\begin{figure}
     \includegraphics[width=.3\linewidth]{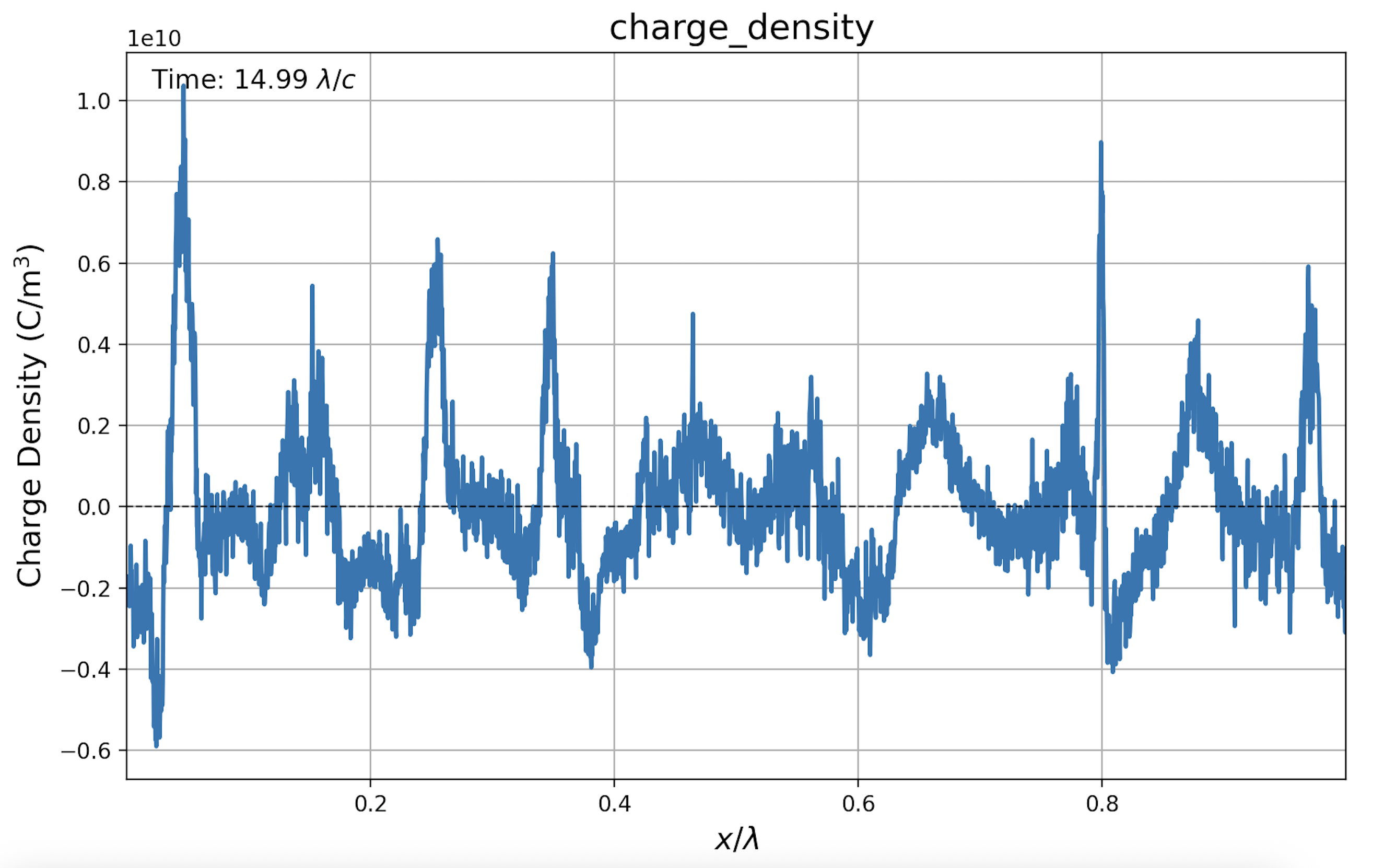}
        \includegraphics[width=.3\linewidth]{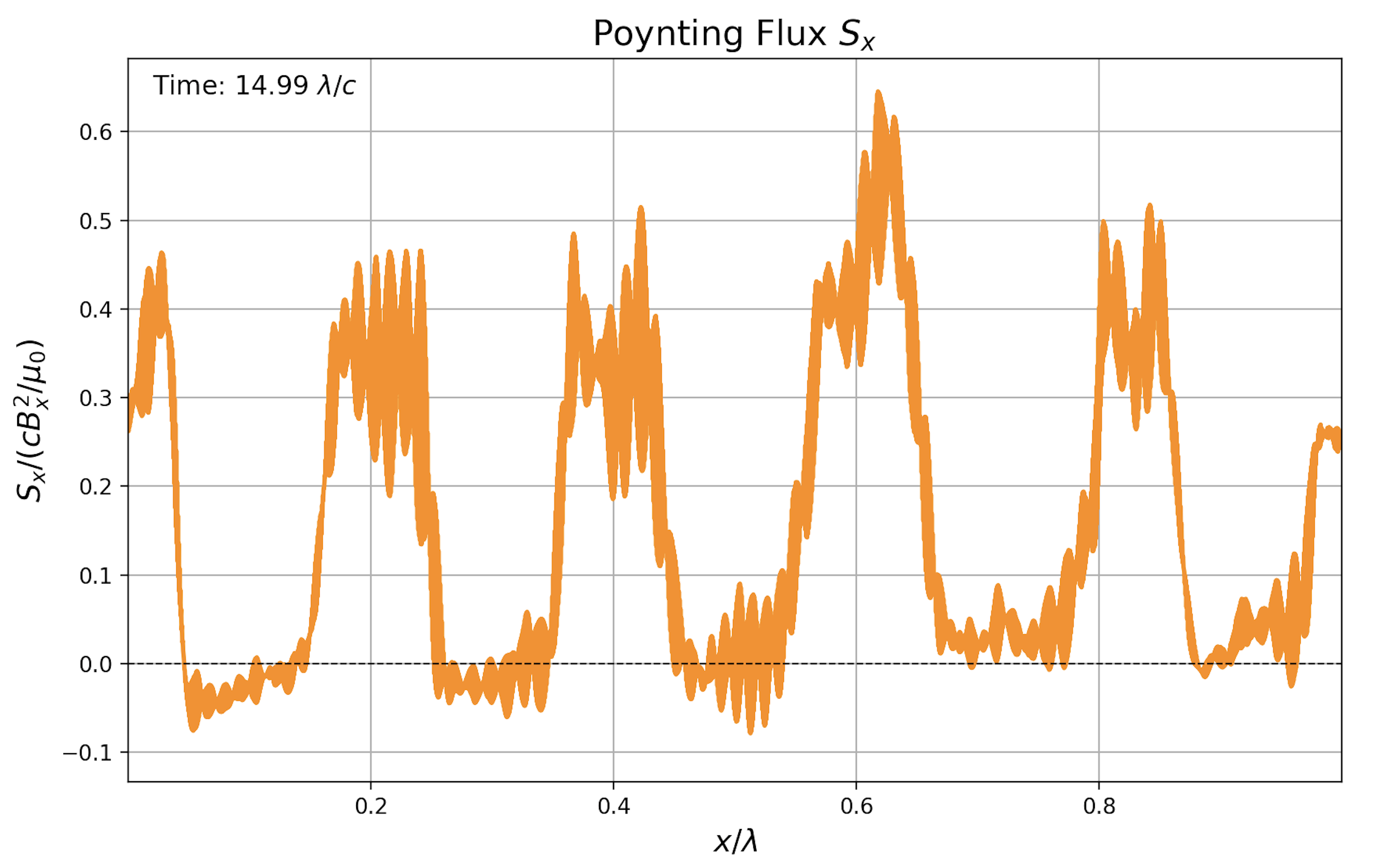}
          \includegraphics[width=.3\linewidth]{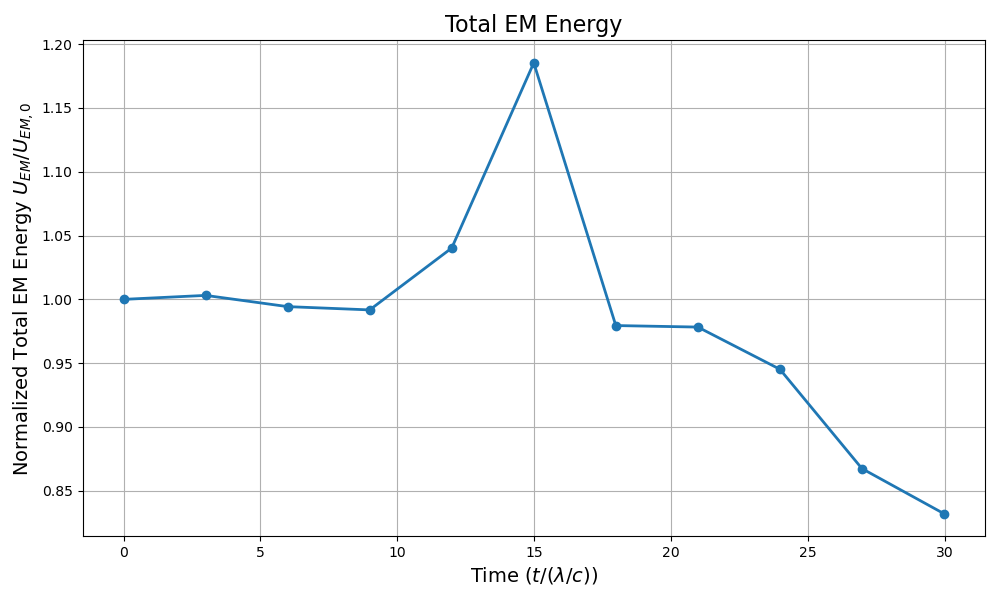}
\caption{ Snapshot of charge density and  Poynting flux  (left and center) and evolution of the EM energy (right). THe snapshots correspond to the time of local peak of \EM\ energy.  }
\label{Langmuir}
\end{figure}

Occasionally we observe transient periods of large charge density oscillations, accompanied by generation of plasma waves, Fig. \ref{Langmuir}. These periods are brief, lead to temporal increase in the EM energy. These are highly dissipative transient phenomena. It is interesting that even in the symmetric  pair plasma one can still produce large charge oscillations. The reason is that for a particular choice of polarization, two signs of charge experience different ponderomotive forces induced by the newly generated  sub-harmonics. Different ponderomotive forces lead to excitation of plasma  wave, that are then quickly dissipated.

\section{Lab frame}

Dispersion of nonlinear  CP\Alfven waves in lab frame were calculated in \cite{2025arXiv250917245L}. For convenience,  the non-linearity parameter was introduced in terms  of fluctuating  \Ef. $\delta _E = E_w/B_0$. Since in the \Alfven frame \Ef\ is zero, we parameterized the fluctuating \Bf\ in lab frame as $\delta_0 =  B_w/B_0$, Then for the \Ef\ we have
$E_w = v_A B_w$ ($ v_A$ depends on amplitude $\delta_0$. 

We find (by analogy with  Eq. (\ref{main}),
\ba &&
\frac{\delta_0  {\cal K} \sigma_0 }{\gamma _A^2}+v_e-v_p=0
\nn && 
v_A \left(\delta_0 +  {\cal K}  p_p\right)-v_p =0
\nn && 
v_A \left( {\cal K} 
   p_e-\delta_0 \right)+v_e =0 
   \nn &&
    {\cal K}  = \frac{k}{\om_B}
    \nn &&
  \sigma_0 = \frac{\om_B^2}{  \om_{p,0}^2}
   \ea
   All quantities are expressed in terms of lab values; definitions of cyclotron $\om_B$ and plasma $\om_{p,0}$ frequencies do not include relativistic corrections to the  effective 
   mass. The  gyration centers of particles are assumed to be at rest in lab  frame (ponderomotive  effects are neglected.
   
   In lab frame the  critical  dimensionless wave number  
   \be
    {\cal K} _0 \equiv \frac{1}{\delta_0 \sigma_0}
    \ee
    which translates to 
    \be
    k_0 =\frac{\omega _p^2}{\delta _0  \omega _B} \equiv \frac{\omega _B}{\delta _0 \sigma_0 }
    \label{k0}
    \ee
    Equation (\ref{k0}) gives a  critical wave-vector of \Alfven waves as expressed in terms of lab quantities: nonlinear \Alfven waves exist only for $k < k_0$.
    
    Importantly, in highly magnetized plasma $\sigma_0 \gg 1$,  for mildly nonlinear waves with $\delta _0 \sim 1$, the critical wave number is $\ll \om_B$.
    
    We also give here an expansion near $k \to k_0$:
    \ba && 
    \epsilon =1-  {\cal K}    \delta  _0 \sigma_0  = 1 - \frac{  {\cal K} }{ {\cal K} _0}
    \nn &&
    p_p = \frac{  \delta  _0^2 \sigma_0 }{\epsilon}
    \nn && 
     p_e= \epsilon
     \nn &&
     p_A = \frac{\epsilon}{  \delta  _0}
     \ea
     
     \begin{figure}
 \includegraphics[width=.99\linewidth]{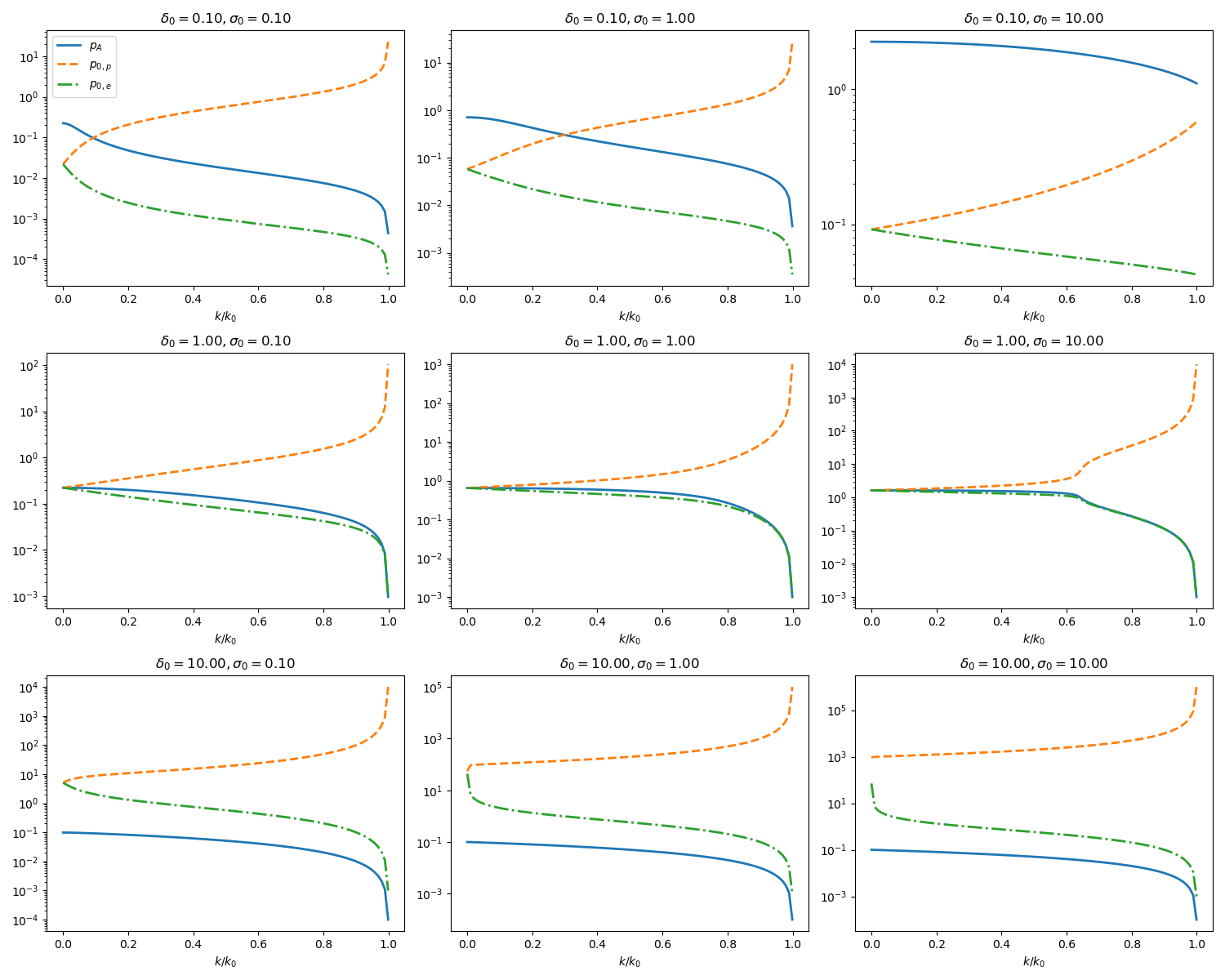}
  \caption{Same as  Fig. \ref{Roots_Tracking}, but in terms of lab frame quantities. }
\label{Roots_Lab_Tracking}
\end{figure}

\begin{figure}
 \includegraphics[width=.99\linewidth]{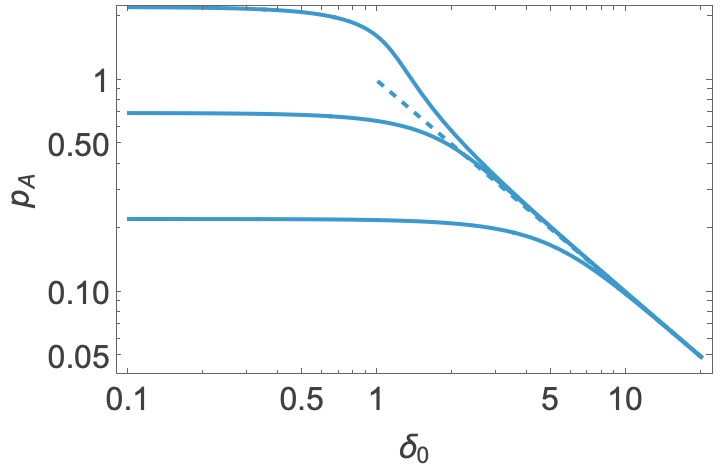}
  \caption{Nonlinear \Alfven momentum $p_A$ in the limit $k \to 0$ as function of non-linearity parameter $\delta_0$ for $\sigma = 0.1, 1 ,10$ (the $\delta_ 0 \to 0$ limit  is $p_A = \sqrt{ \sigma_0/2}$. For large   $\delta_0$, $  p_A \approx 1/ \delta_0$ (dashed line). }
\label{pAofdelta0}
\end{figure}

The $k\to 0$ limit reads
\ba && 
 p_{e,p} = p_0 \mp    \Delta_0    {\cal K} 
 \nn && 
p_0=\frac{\delta_0   p_A}{\sqrt{1+\left(1-\delta_0  ^2\right) p_A^2}} = \frac{\delta_0   \sigma_0 }{2 p_A \sqrt{1+ p_A^2}}
\nn &&
\Delta _0=\frac{\delta  \left(1+ p_0^2\right){}^{3/2} \sigma }{2 \left(1+p_A^2\right)}
\nn && 
\delta_0  ^2=\left(1+ p_A^2\right) \left(\frac{1}{p_A^2}-\frac{4 p_A^2}{\sigma_0 ^2}\right)
\ea
Fig. \ref{pAofdelta0}.
In the linear case $ \delta_0  \to 0$,  we recover conventional $p_A = \sqrt{\sigma_0/2}$  

\begin{figure}
   \includegraphics[width=.49\linewidth]{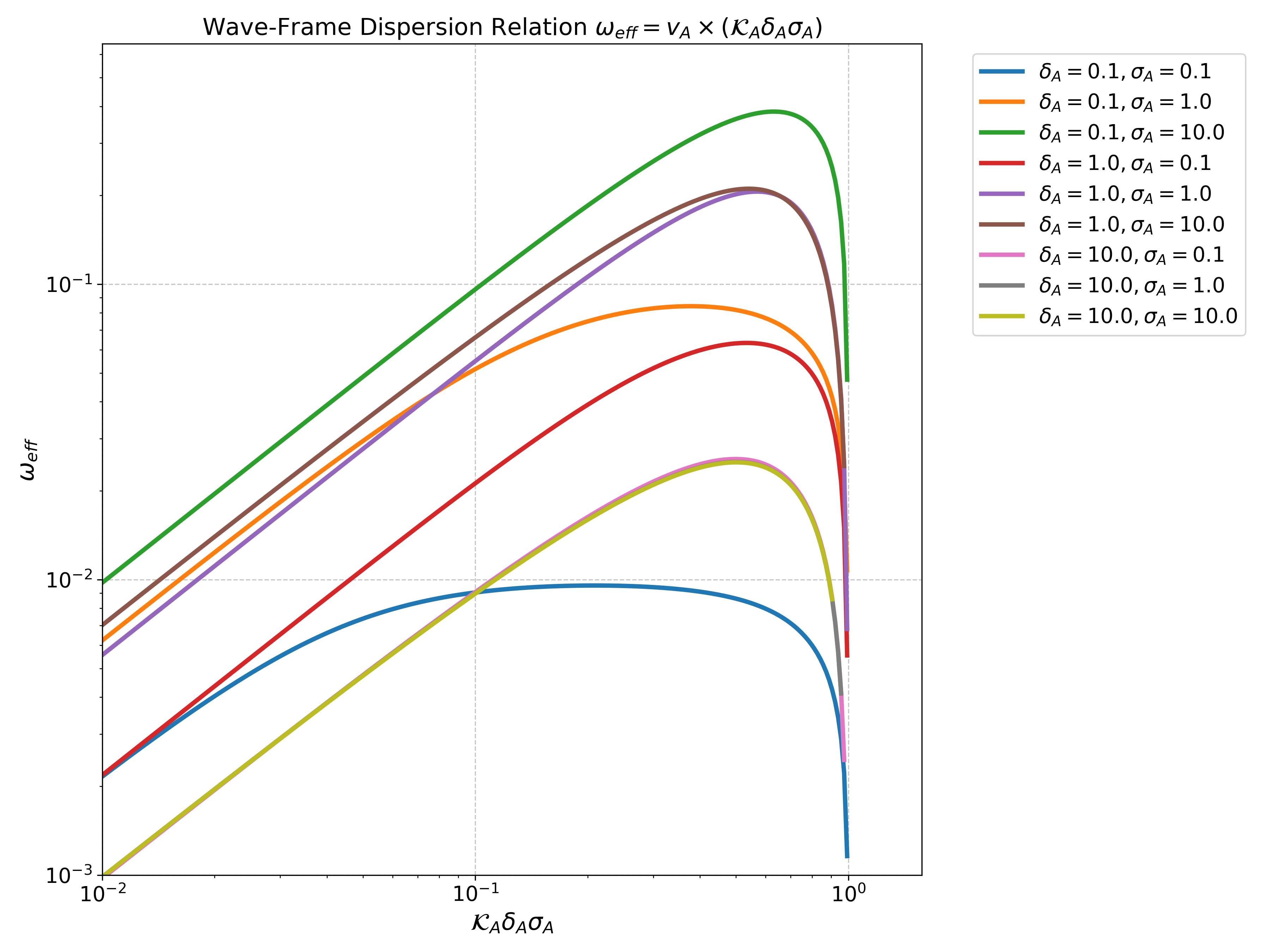}
 \includegraphics[width=.49\linewidth]{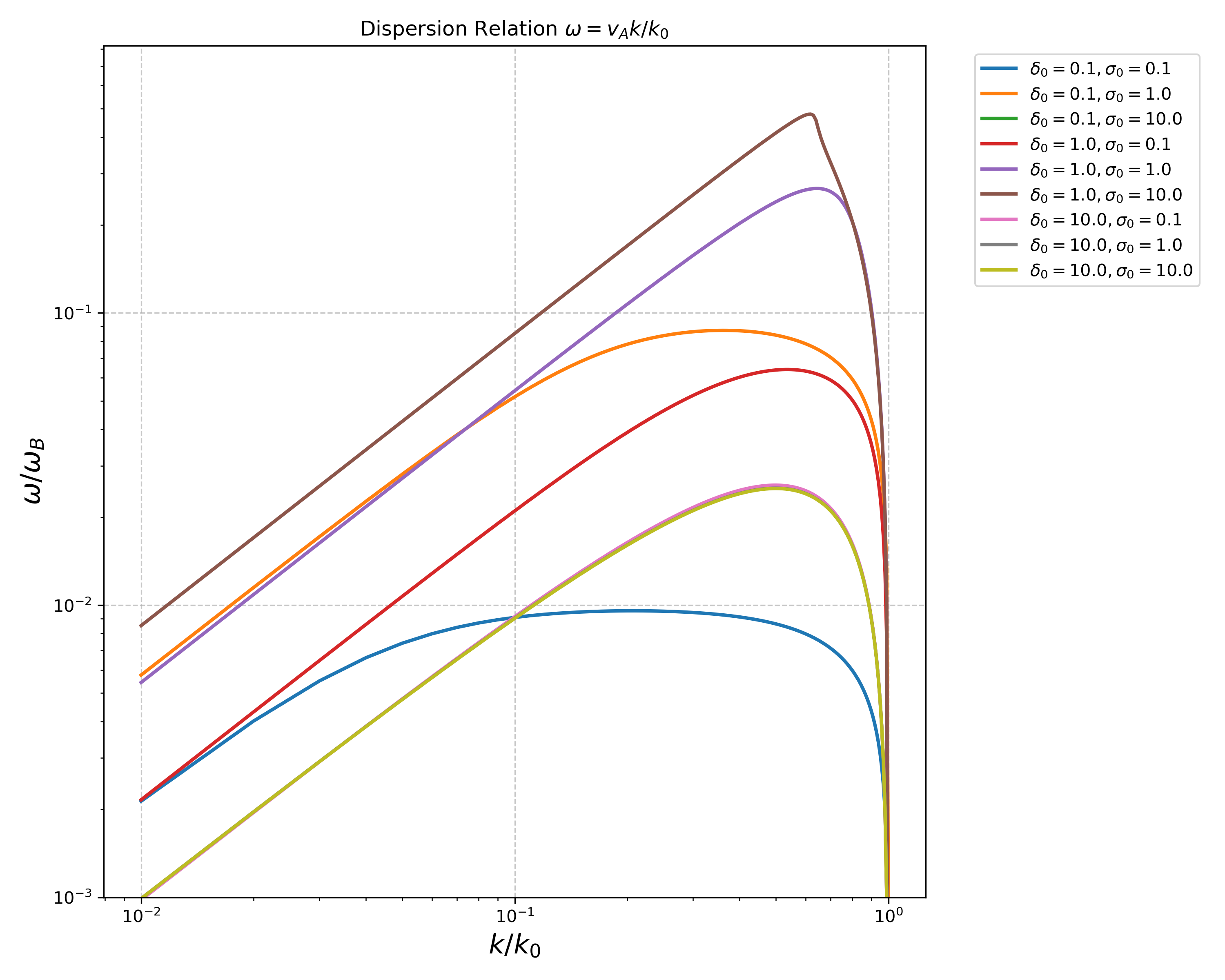}
  \caption{  Dispersion curves   for nonlinear \Alfven waves: effective $\om_{eff} = v_A {\cal K}_A$ in \Alfven frame (left panel), $\om (k, \delta_0)$  (right panel).  Compare with Fig. 7 of \cite{2025arXiv250917245L}.}
\label{domdk}
\end{figure}

Finally, all the dispersion curves show parts with negative group velocity, $\partial \om/ \partial k<0$, Fig. \ref{domdk}. All  simulations  correspond to the stable part  $\partial \om/ \partial k > 0$.

(To add to mathematical complexity,  for low $\delta_0$ regimes, the solution space actually fractures into two  disconnected mathematical manifolds: one starts continuously from $k=0$, then structurally detaches and   passes right past $k = k_0$, another starts at $k/k_0=1$, $\om=0$ and  breaks down  at $k/k_0 \approx 0.989$.)

     \section{Analytical estimates of the parametric instability  in  magnetized pair plasma}

Analytical treatment of parametric instability of weakly nonlinear   waves in  magnetized pair plasma is complicated, but there is  short-cut  in the low frequency limit as we discuss next. Our approach generally follows the so-called  high gain Compton regime of SASE FELs \citep
{1950BSTJ...29..608P,1985NIMPA.239....1S,Freund2024}. 

We  followed  \cite{2025arXiv250920594L} Section III.B. In the linear regime, the corresponding  relations  are easier to derive in the plasma, not \Alfven frame (as  an order of magnitude estimate, relations will be valid for $\sigma \sim 1$ and $\delta_A \leq 1 $

Importantly, for CP waves in  pair plasma in \Bf\ there are charge density oscillations due to the beat between the driver and the backscattered wave. In the small frequency limit,  we can assume that the resulting charge density oscillations are the fastest.  On  longer time scales then the  plasma is charge neutral.  In symmetric pair plasma, charge effect will be proportional to  $\propto 1/\om_B$, while charge-neutral one will be $\propto 1/\om_B^2$.

Following  \cite{2025arXiv250920594L} Section III.B, the evolution of the perturbed amplitude $a_1$ follows from
\ba &&
\partial_ t a_1 = i \delta_0 \frac{\om_{p,0}^2}{\om_B} f
\nn &&
\partial_ t f = - 2 i k_0 \beta
\nn &&
 \partial_ t \beta = 2 i   \delta_0 \frac{k_0 \om_0}{\om_B} a_1
 \label{a1}
 \ea
(since $\delta_0 \propto 1/\om$, the  equations  (\ref{a1}) are in fact  $\propto  1/\om_B^2$).

 We  then find the growth rate   
   \ba &&
{ \Gamma_B }\approx  \rho_B   {\om}_0 
    \nn &&
     \rho_B = \left( \delta_0  \frac{\om_{p,0} }{\om_B} \right) ^{2/3} \approx \delta_0^{2/3} \sigma_0 ^{-1/3}
    \label{GammaA}
\ea
Origin of $ \rho_B$ can be traced to the Pierce parameter  as evaluated in the beam frame \citep{1950BSTJ...29..608P}, and for highly magnetized plasma 
\citep[Comparing parameter $\rho_L$ in Ref. ][with $\rho_B$, one just substitutes $ \omega \to  \omega_B $ to go from  $\rho_L$ to $\rho_B$ ]{2025arXiv250920594L}

Our simulations clearly  follow   the expectations  that for  higher relative fluctuations and smaller magnetization $\sigma$ density fluctuations grow faster. Electromagnetic fluctuations grow slower, and thus typically evolve already in the nonlinear regime of density fluctuations.

\section{Application to magnetars and  FRBs}

Estimating plasma frequency as  \citep{tlk} 
\ba &&
\om_p^2 \approx \frac{ \om_B c}{r}
\nn &&
\sigma_0  \approx \frac{r \omega _B}{c}
\ea
($r > R_{NS} $ is distance to the center of a NS of radius $R_{NS}$; we retain here the speed of light $c$; this is an upper limit on plasma density and $\sigma_0$ parameter),
the critical wavelength for \Alfven waves becomes
\be
\lambda =  \frac
 {2 \pi}{k_0 } = 2 \pi r \delta_0 ,
\ee
a fraction of a radius. (All shorter wavelength will experience modulation.)

 After times 
 \be
 \tau_B \sim \frac{ 1}{\Gamma_B}  = \frac{ 1}{\om_0 \rho_B} =  \left( \frac{r \omega _B}{c} \right)^{1/3} \delta_0^{-2/3}  \omega_0^{-1} 
 \ee
  the dominant wave mode is 
\be
\lambda ' \approx \frac{\lambda}{\sigma_0 } \ll \lambda \sim 2\pi  \delta_0 \frac{c}{\om_B},
\ee
much shorter.

Such extreme transformation of waves  (from macroscopic scale $\lambda \sim r $ to  Larmor scale) clearly cannot occur  - this is a limitation  of our  order-of-magnitude  estimate, where we used parameters for the maximal growth rate. But they indicate that fast generation of shorter waves due to modulational instability of \Alfven\ waves in pair plasma of magnetar \mss\  is possible.

\section{Discusion}

We find that in pair plasma nonlinear \Alfven waves are subject to powerful parametric  modulational  instability.
The  instability is driven by the density fluctuations that  are produced by   a parametrically  unstable  back-scattered wave \citep{2025arXiv250906230T,2025arXiv250920594L,2025arXiv250917245L}. This process is similar to the operation of Free Electron Lasers (FEL)  in the  {\it high gain Compton regime} \citep
{1950BSTJ...29..608P,1985NIMPA.239....1S,Freund2024}.

The most powerful instability occurs near the ``end of the dispersion'',  for wave vectors $ k \leq k_0$ (Eq. (\ref{k0}), in lab frame). Near $ k \approx  k_0$, the instability is exceptionally fast for $\sigma \gg 1$; all modes $k <k_0$ are, formally, unstable.

Unlike the unmagnetized pair plasma case \citep{2025arXiv250917245L}, the backscattered CP wave in magnetized pair plasma does produce axial  {\it charge} density fluctuations.  This adds to the overall  complicated  nonlinear interactions.  We do observe transit excitation of powerful Langmuir waves. But in most regimes, plasma oscillations  are quickly damped, so that the dominated dynamics is charge neutral. 

The key point, similar to \citep{2025arXiv250917245L}, is that density fluctuation in pair  plasma may grow  highly nonlinear. As a result this creates randomly fluctuating perturbations of the permittivity, and the associated Anderson-like localization of the waves. 

On the other hand, the processes may be seen as a type of Weibel instability:  azimuthal  currents associated with the \Alfven wave break-up - this leads to axial  modulation \citep[see Fig. 14 in][]{2025arXiv250920594L}.

Importantly, we do not launch a wave into plasma from outside (a mathematical  boundary value problem), but set it up right inside the plasma (a mathematical eigenvalue problem). The latter case is more relevant for astrophysical set-up, where typically emission is generated by   selected particles already  inside plasma. 

Our simulations, with periodic boundary conditions, likely over-estimate the modulational growth rate, as fluctuations of the density and \EM\ fields are quickly -  and most importantly,    coherently  - ``recycled''. This is a natural  limitation of numerical simulations.

      Two important effects specific to pair plasma  are: 
   \begin{itemize}
    \item  In unmagnetized plasma,  nonlinear electromagnetic waves  experience Anderson self-localization \cite{2025arXiv250920594L}. The beat between the driver and a back-scattered wave creates  charge-neutral,    large  random  density fluctuations  $\delta n/n_0  \gg 1$, and  corresponding  fluctuations of the dielectric permittivity $\epsilon$   (random plasma density grating).  In plasma with guide field, these effects are less pronounced for superluminal waves, but, as we demonstrate here, are important for subluminal  \Alfven waves. 
\item For \Alfven\ waves the  stable part of the dispersion relation $\om(k)$  effectively terminate at finite  $\om^\ast  - k^\ast$,  where  the group velocity becomes zero.  Qualitatively,  subluminal modes with fluctuating \Ef\ larger than the guide field,  $E_w (\om)  \geq B_0$, cannot propagate. 
\end {itemize}

Our results have important application to the plasmas in pulsars and magnetars. For example,  large active regions in \mss\ of \NSs, eg $\sim 1$ km near the surface will initially produce \Alfven waves of similar wavelength.
Modulational instability will  lead to  the generation of much shorter waves, $\sim $ meters. Compton up-scattering by mildly relativistic beam, with $\gamma \sim 10$ in the regime of (magnetized) Free Electron Laser then leads to the generation of coherent  emission \citep{2021ApJ...922..166L,2025JPlPh..91E..10L}.

Second, modulational instability  of \Alfven waves in high-$\sigma$ plasmas may  occur faster than an inertial  turbulent cascade. Also, large density fluctuations will be generated, even by initially incompressible \Alfven waves. 


I would like to thank Xinyu Li, Anatoly Spitkovsky  and Jonathan Zrake for discussions.


\bibliographystyle{apsrev}
\bibliography{/Users/lyutikov/Library/CloudStorage/Dropbox/Research/BibTex}

\appendix

\section{Comment on modulational instability  of waves in pair plasma }
\label{modulational}

Modulational instabilities \citep{LIGHTHILL,1965JFM....22..273W,2009PhyD..238..540Z}  are typically  treated as expansion in small nonlinearity parameter, resulting  in the nonlinear Schroedinger equation. The result is the  Benjamin-Feir-Lighthill criterion for modulational (in)stability. 
In case of pair plasma, even in the simplest case of  CP wave propagating in an unmagnetized  plasma, the analytical approach  to modulational instability  is complicated, riddled with errors. In Ref.
\cite{1983Ap&SS..97....8C}  axial motion was neglected; in Ref. \cite{1984PPCF...26.1099M}  the error was pointed out: the neglect of parallel motion. In Ref. \cite{1989JPlPh..42..507K}, doing apparently correct analysis with the nonlinear Schroedinger equation (NLSE) approach, it was concluded that the CP wave propagating in unmagnetized pair plasma is modulationally stable. 
The problem was addressed in modern PIC simulations in Refs. \cite{2016PhRvL.116a5004E,2017PhRvE..96e3204S}.

Our results, also \cite{2025arXiv250917245L},  contradict the conclusion of stability of nonlinear waves in pair plasma. In fact, the whole modulational instability/NLSE/quasilinear/wave-wave interaction approaches seem not applicable to  nonlinear waves in pair plasma. Typically approach is   an expansion in powers of the  nonlinearity parameter of the wave.   The resulting equations  describe a slow change in amplitude; all parameters (density, slowly varying amplitude) are expanded in a series of nonlinearity. 
This approach  turns out to be inapplicable even to mildly nonlinear waves in pair plasma.  In this case,  instead of expansion by small fluctuations,  quickly   large localized density fluctuations appear. This is the regime of Anderson localization/powerful modulational instability.

\end{document}